\documentclass{aa}

\PassOptionsToPackage{colorlinks=true,allcolors=blue}{hyperref}
\usepackage[varg]{txfonts}
\usepackage{amsmath}
\usepackage{mathtools}
\usepackage{graphicx}
\usepackage{xcolor}
\usepackage{array}
\usepackage{gensymb}  % for \degree symbol in math
\usepackage{tablefootnote}
\usepackage{lipsum}
\usepackage{simples-matrices}  % for vector columns and matrices
\usepackage{comment}

\usepackage{orcidlink}
\usepackage{hyperref}

\begin{document}

\title{
  Warps survive beyond fly-by encounters in protoplanetary disks
}

\subtitle{RW~Aur~A as a case study}

\author{
	C. N. Kimmig\inst{1,2} \orcidlink{0000-0001-9071-1508}
    \and P. Weber\inst{3,4,5} \orcidlink{0000-0002-3354-6654}
    \and G. P. Rosotti\inst{1} \orcidlink{0000-0003-4853-5736}
    \and S. Facchini\inst{1}  \orcidlink{0000-0003-4689-2684}
    \and C. P. Dullemond\inst{2} \orcidlink{0000-0002-7078-5910}
}

\institute{
    Dipartimento di Fisica, Università degli Studi di Milano, via Celoria 16, 20133 Milano, Italy 
    \and
	Zentrum f\"ur Astronomie, Heidelberg University, Albert-Ueberle-Str.~2, 69120 Heidelberg, Germany
	\and
	Departamento de Física, Universidad de Santiago de Chile, Av. V\'ictor Jara 3493, Santiago, Chile \and
        Millennium Nucleus on Young Exoplanets and their Moons (YEMS), Chile \and
Center for Interdisciplinary Research in Astrophysics Space Exploration (CIRAS), Universidad de Santiago, Chile
}

\date{Received xxx / Accepted xxx} % format day month year e.g. 1 January 2000

\abstract
{}
{Stellar fly-bys can have multiple dynamical effects on protoplanetary disks, including warping and the excitation of spiral arms. Since observations indicate that warps are common, we aim to investigate these effects for different fly-by trajectories.
We further link our models to observations by applying them to the RW~Aur system, which is a fly-by candidate with a relatively well constrained trajectory.
}
{We investigate the disk dynamics in grid-based hydrodynamical simulations, which allow for a lower disk viscosity than commonly used SPH models.
We post-process our simulations of the RW~Aur system with radiative transfer models to create synthetic images of the dust continuum and gas kinematics.
}
{Fly-bys inclined with respect to the original disk plane can excite warps of a few degrees: the exact outcome depends on the specific geometry of the encounter. Specifically, we find that the position of the periastron with respect to the initial disk plane plays a role for the resulting warp strength. Within our parameter set, the strongest warp is excited for a retrograde fly-by with a periastron that is not in the same plane as the disk.
Our models show that the warp can persist even after the perturber can no longer be clearly linked to the system, implying that past fly-bys are a possible origin of observed warps.
Excited spirals arms, on the other hand, are much more short-lived than the warp.
The RW~Aur system presents a perfect opportunity to apply these results: we find that a warp of about 5\degree can be excited, and that the strong spiral arms have already disappeared at the current time of observation ($300\,\mathrm{yr}$ after periastron). This compares well with existing continuum observations, and our synthetic kinematic evaluations hint at remnant structures in the gas density that may be detectable.}
{}

\keywords{protoplanetary disks -- methods: numerical -- hydrodynamics -- radiative transfer}
\maketitle

\section{Introduction}

Most stars are not born in isolated environments.
Usually, they form in large star-forming regions, often in clusters of many thousand stars \citep{Draine2011, Krause2020}.
Observations show that these star-forming regions are highly dynamic with significant relative velocities between stars \citep{Karnath2019, Kuhn2019, Yang2025}.
Thus, young stars are likely to encounter other young stars \citep[see e.g.][]{Pfalzner2013, Bate2018, Lebreuilly2021, Rawiraswattana2023, Lebreuilly2024}.
These stars commonly host planet-forming disks and therefore the interactions in a stellar cluster can affect the disk shape and morphology, and in turn influence the process of planet formation \citep{Kobayashi2001, Fragner2009, Thies2010, Breslau2019, Cuello2019, Aly2020, Aly2021, Longarini2021}.

Close encounters on unbound orbits are commonly referred to as stellar fly-bys.
Often, a fly-by is defined to have a separation between the two stars of less than $1000\,\mathrm{au}$ \citep{Cuello2023}.
As the occurrence rate of fly-bys is highest while the objects are still spatially confined to their star-forming region, the probability at which disk-hosting stars experience close gravitational interactions with other stars in their lifetime can be as high as $70\%$, depending on the stellar density \citep[][see their Fig.~7]{Winter2018}.
\citet{Cuello2023} conclude in their review, that more than half of young stars hosting a disk experience a fly-by with a separation of less than $1000\,\mathrm{au}$.

Stellar fly-bys have a significant impact on the dynamics and morphology of disks around either or both of the fly-by components.
One of the first numerical studies of the effect of fly-bys on circumstellar disks was performed by \citet{Clarke1993}, and the topic has been investigated extensively since.
These effects include spiral arms due to tidal perturbations \citep{Ostriker1994, Pfalzner2003, Cuello2019, Smallwood2023}, truncation of disk sizes \citep{Breslau2014, Rosotti2014, Munoz2015, Vincke2015, Bhandare2016},
ejection and capture of disk material by the fly-by object \citep{Heller1995, Jilkova2016, Cuello2019, Cuello2020, Smallwood2024}, and misalignment and warping of disks \citep{Terquem1993, Picogna2014, Xiang-Gruess2016, Nealon2020}. 

The resulting shape of the disk depends on the geometry of the fly-by orbit with respect to the disk.
Fly-by trajectories are typically classified as either prograde, retrograde, or polar, depending on the direction of the angular momentum vectors of the orbit and the disk. For prograde fly-bys, both vectors are in the same hemisphere and in retrograde fly-bys in opposite hemispheres. Mutual orientations of these vectors close to $90\degree$ are called polar.
Prograde orbits are found to be most disruptive \citep{Clarke1993, Xiang-Gruess2016, Winter2018}. In these cases, the spirals caused by the tidal torques are strongest, and the truncation of disks is most effective.
For example, \citet{Bhandare2016} find for retrograde orientations that disk sizes after truncation can be twice as large as for prograde fly-bys.
A characteristic signature of these retrograde non-coplanar fly-bys is a warp due to the misaligned gravitational torque \citep{Terquem1993, Xiang-Gruess2016, Nealon2020, Cuello2023}, although disks can also become warped in inclined prograde fly-bys \citep[see e.g.][]{Larwood1997}.

Warps in protoplanetary disks have gained importance throughout recent years, as more and more observations reveal indications that a significant fraction of disks is warped \citep{Ansdell2020, Kluska2020, Bohn2022, Garufi2022, Benisty2023, Winter2025}.
For example, non-axisymmetric shadows can hint at a misaligned inner region of the disk that blocks the light from the star \citep[][and many more]{Marino2015, Benisty2017, Debes2017, Stolker2017, Muro-Arena2020, Villenave2024, Zurlo2024}.
Indications for warps can also be found in kinematic observations (\citealt{Walsh2017, Mayama2018, Phuong2020, Garg2021}; see also \citealt{Pinte2023} for a review).
After initial excitation, the evolution of the warp shape is controlled by internal torques.
These torques arise from pressure gradients and resonant motions that form due to the mutual misalignment of radially adjacent orbits \citep{Ogilvie2013a, Ogilvie2013b, Dullemond2022}.
The evolution of the warp depends on the viscosity and the disk thickness, characterized by the Shakura-Sunyaev parameter $\alpha$ \citep{ShakuraSunyaev1973} and the aspect ratio $h$, respectively. In thick disks with low viscosity ($\alpha < h$), the warp travels in a wave through the disk \citep{Papaloizou1995, Lubow2000, Nixon2016, Martin2019, Kimmig2024}.
In more viscous disks, where $\alpha > h$, the warp evolves diffusively \citep{Papaloizou1983, Pringle1992, Ogilvie1999}. In typical protoplanetary disks, the viscosity is low relative to the aspect ratio, placing them in the wave-like regime.

Most numerical studies investigate the effects of fly-bys utilizing Smoothed Particle Hydrodynamics (SPH) simulations \citep[see][etc.]{Picogna2014, Xiang-Gruess2016, Cuello2019, Menard2020, Nealon2020}, because these simulations do not rely on a specific geometry for the discretization.
However, SPH methods usually need to include an artificial viscosity which is often higher than the physical viscosities found in protoplanetary disks \citep[for example by][]{Villenave2024}.
In light of the recent development \citep{Rabago2023, Rabago2024, Kimmig2024}, we take to opportunity to focus on grid-based hydrodynamic models of low-viscosity disks subjected to a stellar fly-by.
Stellar fly-bys provide an ideal laboratory for investigating the evolution of warps, as they create a physically motivated warp shape instead of a parametrized assumption, and only distort the disk once without further influence of external torques.

As an additional motivation for our work, we aim to explore different orientations of trajectories with respect to the disk plane.
Recent studies mainly investigate orientations where the trajectory is inclined in such a way that the periastron remains in the same plane as the disk \citep{Cuello2019, Nealon2020, Cuello2023}.
This is likely motivated by the fact that \citet{Xiang-Gruess2016} found that for the final mean tilt of the disk, the inclination of the trajectory with respect to the disk is more relevant than the relative position of the periastron.
However, the radial inclination profile could be strongly affected by the position of the periastron. In young stellar clusters, there is no physical constraint that the periastron should be in the same plane as the disk.
In fact, a scenario with the periastron outside the disk plane should be statistically more likely.

In the first part of this work, we therefore revisit a limited set of disk-orbit orientations to compare the effects on the warping of the disk.
We describe our numerical setup in Section~\ref{sec:numerical-method} and present the results of the hydrodynamical investigations in Section~\ref{sec:parameter-study}. In the second part, we apply the models to the observed RW~Aur system and present synthetic observations in Section~\ref{sec:rwaur}. We discuss our work in Section~\ref{sec:discussion} and conclude in Section~\ref{sec:conclusion}.

\section{Numerical method} \label{sec:numerical-method}

\subsection{Hydrodynamic simulations}

If a nearby star approaches a disk close enough, its gravitational pull begins to notably impact the dynamics of the disk. In cases where the fly-by trajectory is inclined relative to the disk plane, this interaction can generate a warp.
To investigate the dynamical effect of fly-bys on low-viscosity disks, we perform three-dimensional hydrodynamic simulations using the grid-based code FARGO3D \citep{FARGO3D, Masset2000}, where we include a gravitational perturber on a parabolic orbit modeled with the built-in orbit solver.

We run gas-only simulations and do not make use of the multi-fluid or magnetohydrodynamic features of the code. For the viscosity, we use an $\alpha$-viscosity model \citep{ShakuraSunyaev1973} with the dimensionless parameter ${\alpha = 10^{-3}}$.
The capability of the code to model warps and disk planes inclined with respect to the grid geometry was extensively tested in \citet{Kimmig2024}.

We model the hydrodynamics on a spherical grid ($r$, $\theta$, $\phi$). Radially, the grid extends from ${2.6\,\mathrm{au}}$ to ${41.6\,\mathrm{au}}$ with $120$ logarithmically spaced grid cells. In the azimuthal direction, we use $100$ cells.
Vertically, we cap off the poles in order to save computation time and include a range of ${\theta_\mathrm{min} = 44.2\degree}$ to ${\theta_\mathrm{max} = 135.8\degree}$ (where $\theta = 90\degree$ corresponds to the initial disk midplane) with $256$ vertical grid cells.

We set up an initially planar (unwarped) disk, which is aligned with the grid midplane.
To ensure the disk remains unaffected by the outer radial boundary, we taper its density profile so that it lies well within the computational domain.
We therefore adopt for the initial surface density $\Sigma$ profile
\begin{equation} \label{eq:surfdens}
\Sigma(r) = \Sigma_0 \left(\frac{r}{r_0} \right)^{-p} \left[1 + \exp \left(\frac{r - r_\mathrm{out}}{0.05 r_\mathrm{out}}\right) \right]^{-1}
\left[ {1 + \exp\left(\frac{r_\mathrm{in} - r}{0.05 r_\mathrm{in}}\right)} \right]^{-1},
\end{equation}
where $\Sigma_0$ is the surface density at the reference radius $r_0$, for which we use ${r_0=5.2\,\mathrm{au}}$.
The equation includes exponential cut-offs at both disk edges, where $r_\mathrm{in}$ and $r_\mathrm{out}$ are the inner and outer edge of the disk, respectively. Unless specified otherwise, we use ${r_\mathrm{in}=3.12\,\mathrm{au}}$ to ${r_\mathrm{out}=26\,\mathrm{au}}$, a slope of $p = 1$, and a total disk mass of ${M_\mathrm{disk}=0.02\,M_\odot}$, which leads to a surface density ${\Sigma_0=209\,\mathrm{g/cm^2}}$ at $r_0$.
Since the disk-to-star mass ratio lies below $0.5$, gravitational instability is not expected to be relevant.
Earlier work demonstrates that self-gravity can modify the warp evolution and damping in low-viscosity disks \citep{Papaloizou1995}. Yet, wave-like warp propagation is also expected when self-gravity is negligible \citep{PapaloizouTerquem1995}. To avoid additional complexity, we neglect self-gravity in our models.
To vertically expand the surface density, we use the default setup of the example \texttt{p3disof} from FARGO3D.

We assume a locally isothermal model, where we set the temperature structure so that the aspect ratio $h$ takes the form
\begin{equation} \label{eq:aspectratio}
h(r) = h_0 \left(\frac{r}{r_0} \right)^{f},
\end{equation}
with $h_0$ being the aspect ratio at $r_0$ and the flaring index $f$. For all our models, we use the values $h_0~=~0.05$ and $f = 0.25$.
The initial azimuthal velocity is set to the Keplerian velocity with a correction for pressure gradients, as described in Appendix~\ref{sec:azimuthalvelo}.

In all simulations, we adopt outflow boundaries for the radial direction, where we allow mass to leave the grid domain, but not to enter.
This means that the radial velocity is copied from the active domain edge to the ghost cells if positive, but forced to zero if negative.
This helps to avoid unphysical influx of gas, which has sometimes been found to occur in fly-by scenarios especially at the inner boundary.
We additionally make use of the wave-killing implementation in FARGO3D based on \citet{Val-Borro2006} and use a width of 10\% of the edge radius for our damping zones.
Vertically, we set reflective boundaries and azimuthally periodic conditions.

We set the grid origin to the central star hosting the disk. Therefore, we need to account for non-inertial motion of the reference frame due to the fly-by. This is taken care of with the build-in indirect term in FARGO3D.
However, we do not treat any non-inertial effects imposed by a possible acceleration from the disk onto the star, for example due to asymmetries in the disk.
As disks in fly-bys can become very asymmetric due to the spirals and the warping, this could potentially have an effect on the dynamics.
However, the detailed physics and the correct implementation of this indirect term are still under investigation \citep{Crida2025, Jordan2025}, and we decided to neglect this effect for now.

\subsection{Fly-by geometry}

\begin{figure}[ht!]
    \centering
    \includegraphics[width=0.8\linewidth]{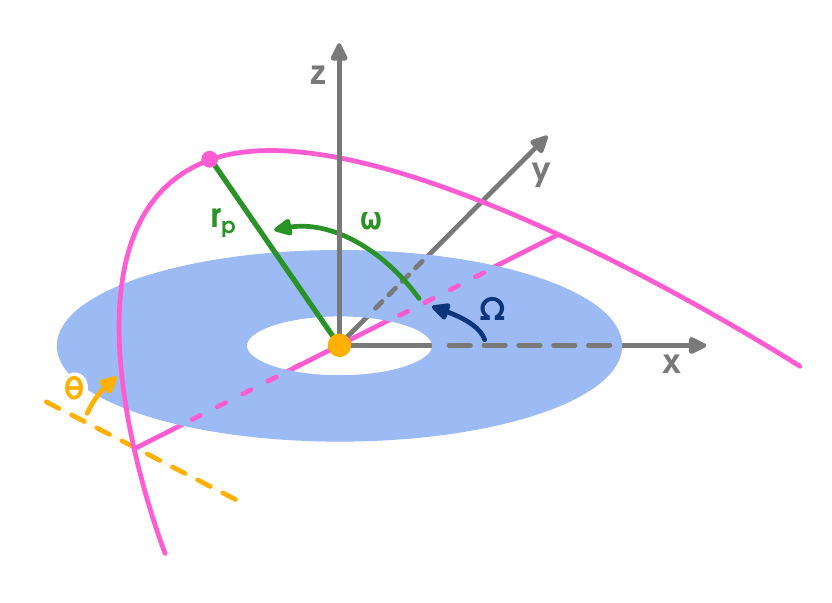}
    \caption{Schematics of the definition for the geometry of the fly-by trajectory with respect to the disk plane. Indicated are the inclination of the trajectory $\theta$, the longitude of ascending node $\Omega$, and the argument of periapsis $\omega$. The distance of the closest approach, i.e. the periapsis, is $r_\mathrm{p}$.}
    \label{fig:flyby-geometry}
\end{figure}

The trajectory of the fly-by is described in Cartesian coordinates. We define the $x$-$y$-plane to coincide with the initial disk midplane.
A convenient way to describe the trajectory orientation are the orbital elements, which are the three angles indicated in Figure~\ref{fig:flyby-geometry}.
We call the inclination between trajectory and disk plane $\theta$.
The longitude of ascending node $\Omega$ is the angle between the $x$-axis and the intersection line between the plane of the trajectory within the $x$-$y$-plane, and the argument of periapsis $\omega$ is the angle between the intersection line and the connecting line of origin to periapsis.
We note that physically, the longitude of ascending node $\Omega$ does not make a difference in our simulations, as the disk is initially axisymmetric, which means that the setup would lead to the same result for different $\Omega$, only rotated.

The trajectory of the fly-by is calculated by FARGO3D with a fifth-order Runge-Kutta N-body solver and we only set the initial position and velocity of the star on the fly-by trajectory, called perturber from here on.
In addition to the orbital elements, the trajectory is characterized by the distance of periapsis (or closest approach) $r_\mathrm{p}$ and the orbit eccentricity $e$.
For the simulations, we choose the initial distance $d_\mathrm{ini}$ between the two stars and then calculate the initial position and velocity of the perturber using the hyperbolic equation. More details are given in Appendix~\ref{sec:appendix-traj}.

\section{Parameter exploration} \label{sec:parameter-study}

In this section, we explore a limited set of fly-by configurations.
We aim to investigate the dynamical effects of a fly-by on the disk, where we focus on disk warping.
We do not intend to cover the full parameter space, nor do we intend to maximize the warp strength of the disk,
but instead choose a set of representative test cases.

\begin{figure}[ht!]
    \centering
    \includegraphics[width=\linewidth]{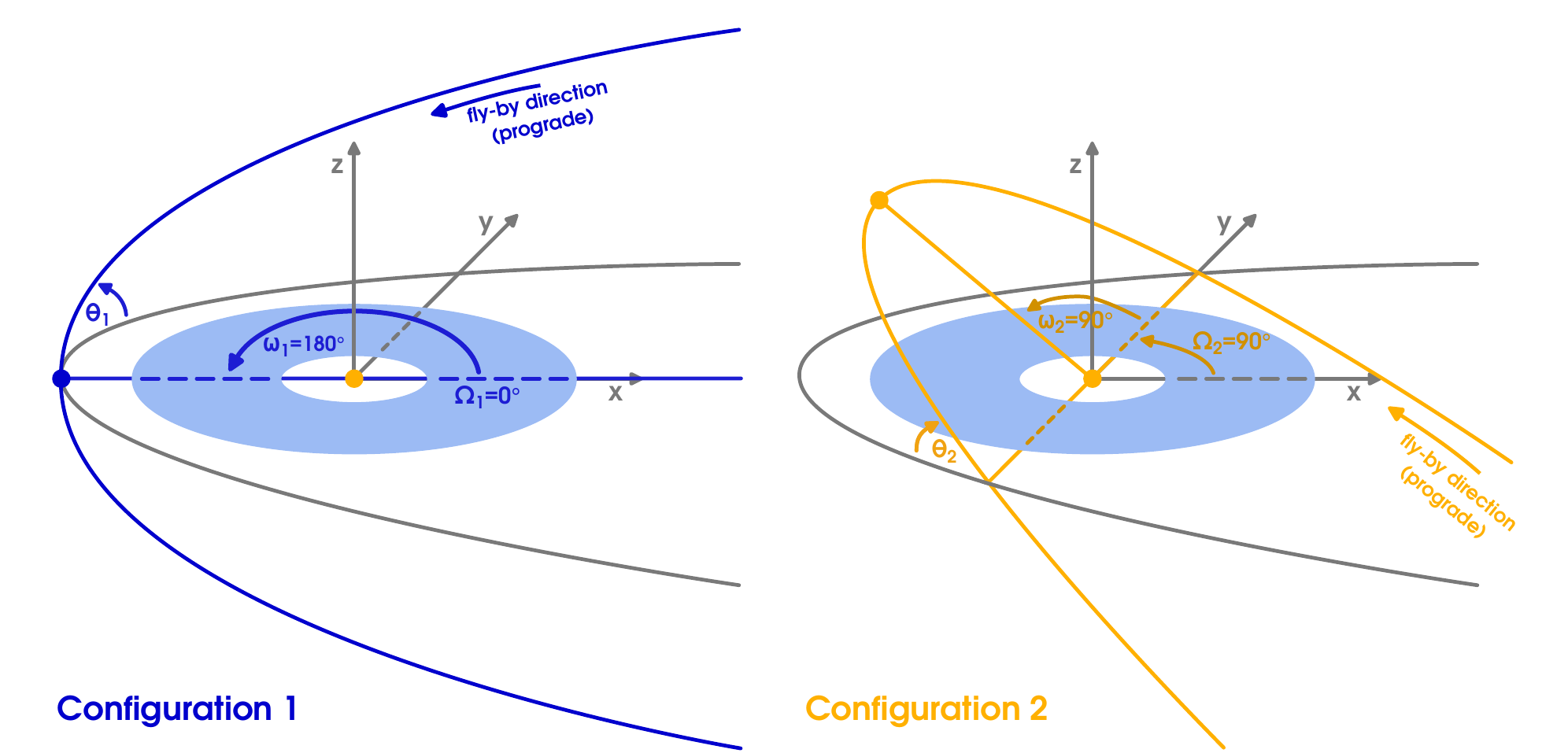}
  \caption{Trajectory configurations for our fly-by simulations. The disk lies in the $x$-$y$-plane at the origin of the coordinate system with a counter-clockwise rotation.
  Configuration~1 is shown in blue on the left, where the trajectory is rotated about the $x$-axis with a periapsis in the same plane as the disk. Configuration~2 (orange, right) is rotated about the $y$-axis with a periapsis out of the disk plane.}
    \label{fig:trajectories}
\end{figure}

Specifically, we decided on two different geometric configurations, depicted in Figure~\ref{fig:trajectories}. For simplicity, we define the trajectories in both configurations such that the perturber comes from the positive $x$-direction, leading to a periapsis in the negative $x$-regime.
For the first configuration, the periapsis lies in the same plane as the disk and the trajectory is rotated about the $x$-axis, corresponding to a longitude of ascending node of $\Omega_1=0^\circ$ and an argument of periapsis $\omega_1=180^\circ$ (see Figure~\ref{fig:flyby-geometry} for the definition of these angles).
The second configuration is rotated about the $y$-axis, and therefore $\Omega_2=90^\circ$ and $\omega_2=90^\circ$.
We set the inclination of the trajectories in both cases to $\theta_{1,2}=30^\circ$.

For both configurations, we perform both a prograde and a retrograde fly-by.
Here we note that technically, the definition of the longitude of ascending node $\Omega$ would change for the retrograde orbit, as the ascending node flips by $180^\circ$. However, for simplicity, we use the same angles to describe both prograde and retrograde orbits and simply flip the starting point of the perturber.
For comparison, we additionally perform a co-planar orbit (with $\Omega=90^\circ$, $\omega=90^\circ$, and $\theta=0^\circ$), also both prograde and retrograde.

Throughout this work, we define $t=0$ to be the moment of the closest approach between the two stars.
We initialize our simulations at $t=-2012\,\mathrm{yr}$, which allows for about $15$ orbits at the outer disk edge ($r=26\,\mathrm{au}$) before the closest approach. %actually 13.6 orbits
Most simulations end at $t=1995\,\mathrm{yr}$, when the perturber has roughly reached the same distance as in the beginning.

In all six simulations, we set the distance of the closest approach at the periapsis to ${r_\mathrm{p}=104\,\mathrm{au}}$.
We model parabolic fly-bys with an eccentricity of $e=1$, and set the initial distance of the perturber from the disk-hosing star to ${d_\mathrm{ini}=1040\,\mathrm{au}}$, so that the disk has enough time to relax from the initial conditions.
All six simulations model an equal mass fly-by, where ${q={M_\mathrm{fl}}/{M_*}=1}$. Here, $M_*$ corresponds to the disk-hosting star and $M_\mathrm{fl}$ to the mass of the perturber. We set ${M_* = M_\mathrm{fl} = 1\,M_\odot}$ for both stars.

\subsection{Disk warping}

We can characterize the warping of the disk by studying the evolution of the inclination profile.
Analogously to \citet{Kimmig2024}, we compute the angular momentum vector of each radial shell of the grid and compute the inclination at each radius as the angle between the local angular momentum vector and the $z$-axis.

\begin{figure} [ht!]
  \centerline{\includegraphics[width=0.35\textwidth]{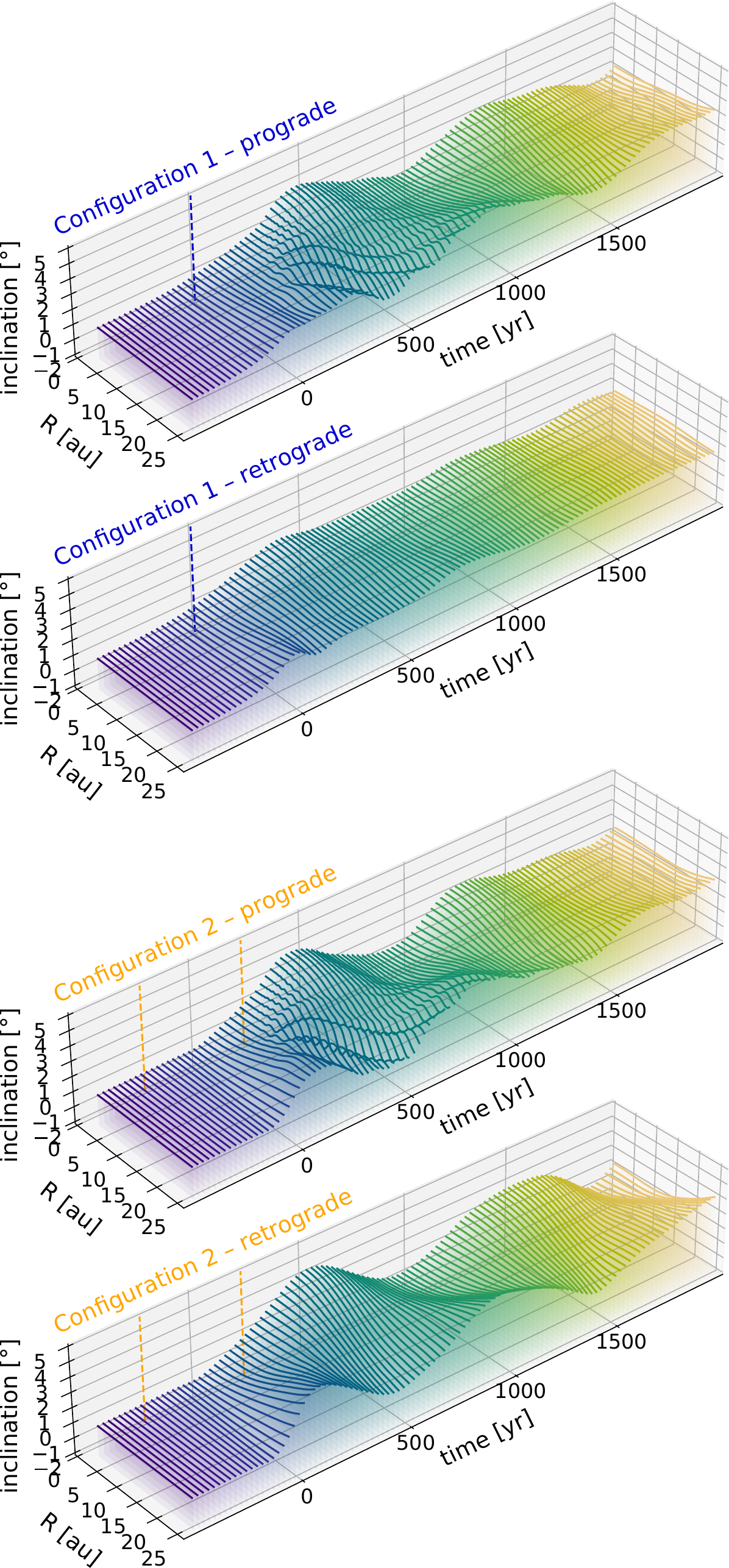}}
  \caption{\label{fig:flyby-incl} Inclination evolution of the fly-by simulations with prograde and retrograde fly-by of Configuration~1 (periastron in the same plane as the disk) in the top two panels and Configuration~2 (periastron out of disk plane) in the bottom 2 panels. Time $t=0$ indicates the moment of closest approach. The color corresponds to time and the vertical blue and orange dashed lines indicate the times when the perturber crosses the (initial) disk midplane.
  We note that we plot the inclination profile only up to the outer radius of the disk $r_\mathrm{out}=26\,\mathrm{au}$, but the computational domain extends further out.
  }
\end{figure}

Figure~\ref{fig:flyby-incl} shows the inclination evolution for both configurations for pro- and retrograde orbits. We checked that the inclination of the disk for a co-planar fly-by does not change, which means that the disk does not warp in these cases, as expected.

Fly-bys on an inclined trajectory, however, all induce a change of the disk inclination with a small warp in most cases. Only the retrograde case for Configuration~1, where the trajectory is tilted about the $x$-axis (but the periastron is in the disk's midplane), inhibits almost no warp. Here, the disk tilts rigidly by about $2^\circ$.
In the other three simulations, the disk becomes warped, and the warp travels in a wave through the disk.
This wave shows the typical standing wave behavior of the global bending modes with a wavelength of $2\,r_\mathrm{out}$.
In addition to the warp, we find a twist, or in other words, a differential precession of the disk. Details are given in Appendix~\ref{sec:appendix-twist}.

Both prograde simulations show features in addition to the warp wave which are excited around the time of closest approach. These features move outward and dissipate quickly and are likely linked to the spiral arms that are excited more strongly in prograde fly-bys.
We will take a closer look at the spiral arms in Section~\ref{sec:spirals}.

\begin{figure}[ht!]
    \centering
    \includegraphics[width=\linewidth]{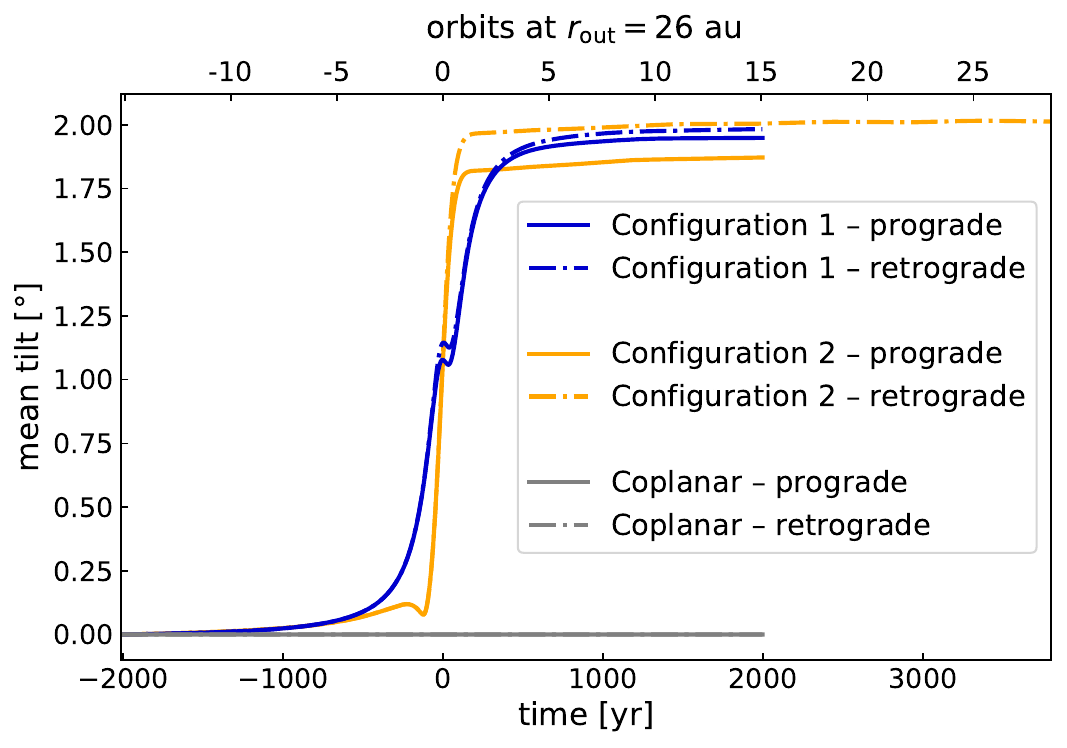}
    \caption{Mean tilt defined as the angle between the total angular momentum vector of the disk and the $z$-axis for all six trajectory configurations.}
    \label{fig:mean-tilt}
\end{figure}

\begin{figure} [ht!]
  \centerline{\includegraphics[width=0.5\textwidth]{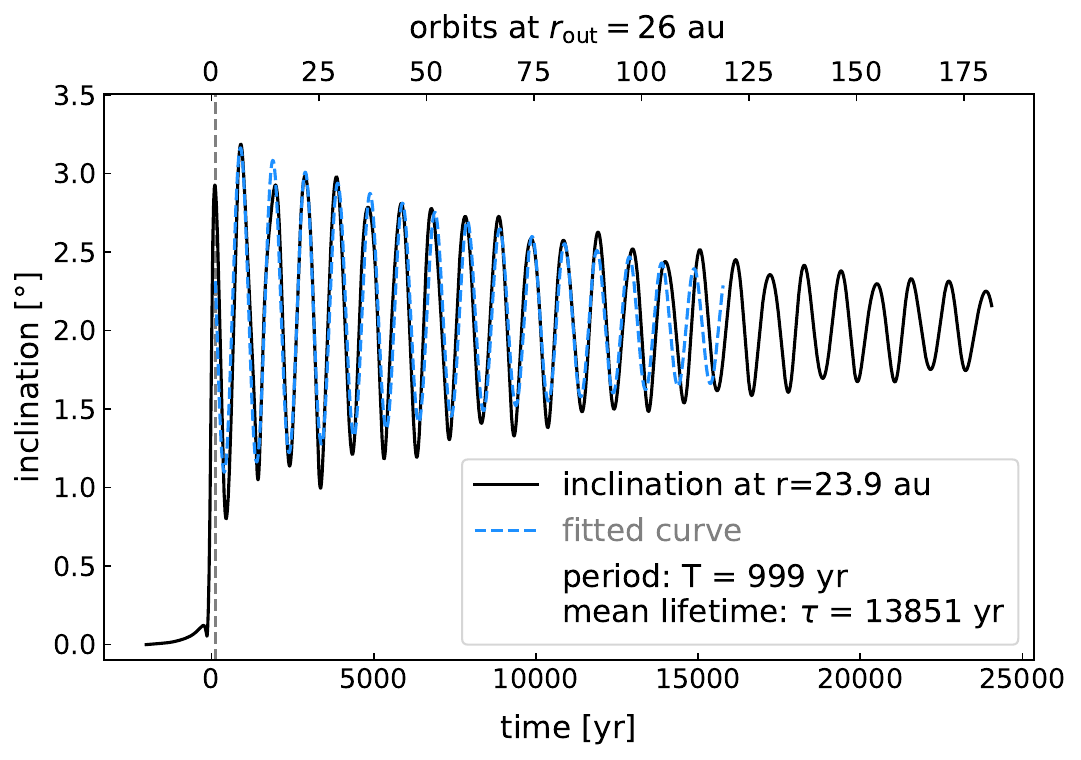}}
  \caption{\label{fig:incl-decay}
  Evolution of the orbital plane at $r=23.9\,\mathrm{au}$ in the retrograde simulation in Configuration~2 (see Figure~\ref{fig:trajectories}, right). We start the fit according to Equation~\ref{eq:fit} (blue dashed line) at the first maximum of the curve (grey dashed line) and stop the fit at $t=1.58 \times 10^4\,\mathrm{yr}$, where the period seems to change. We note that this figure shows the evolution over a longer time than Figure~\ref{fig:flyby-incl}.}
\end{figure}

To investigate the mean tilt of the disk due to the fly-by, we evaluate the mean inclination in each simulation by computing the angle of the total angular momentum vector with respect to the $z$-axis.
Figure~\ref{fig:mean-tilt} shows the evolution of the mean tilt for all simulations.
All inclined fly-bys induce a mean tilt of close to $2^\circ$, irrespective of the exact geometry of the trajectory.
This is consistent with the results by \citet{Xiang-Gruess2016}, who find that the absolute value of the inclination $\theta$ of the orbit with respect to the disk is the decisive quantity determining the final mean tilt of the disk.
Looking at the inclination profiles, we find, however, that the warping of the disk very well depends on the exact orientation of the trajectory.
In particular, the warp is stronger for the retrograde case in Configuration~2, where the periapsis does not lie in the same plane as the disk.
Because the warp is strongest in this simulation (a maximum misalignment of about $4\degree$), we decided to investigate this simulation on longer timescales.
This way, we can study the decay of the warp after the perturber has left the system.
Figure~\ref{fig:incl-decay} shows the evolution of the orbital plane inclination at $r=23.9\,\mathrm{au}$, which is close to the outer edge of the disk.

We can estimate the decay of the warp by fitting
\begin{equation}\label{eq:fit}
f(t) = A \cos(\omega t - b) \times \exp(-\lambda_\mathrm{d} t) + c + d \ t,
\end{equation}
where $A$ is the amplitude parameter, $\omega$ the frequency, $\lambda_\mathrm{d}$ the decay rate, $b$ a possible phase shift, $c$ an offset which corresponds to the mean tilt, and $d$ a linear damping in overall inclination.
The latter parameter is included for possible inclination damping due to the grid, as found in \citet{Kimmig2024}. However, in our simulations with such a low warp amplitude, we find almost no damping, that is, ${d=10^{-5} \deg  \,\mathrm{yr}^{-1}}$.
For the fit, we use the \texttt{scipy}-function \texttt{curve\_fit} \citep{SciPy2020}.

We find that some level of warping is still present after the perturber is already long gone.
The mean lifetime of the warp is around ${\tau = 1/\lambda_\mathrm{d} \approx 1.4 \times 10^4\,\mathrm{yr}}$ and the period of the warp wave about $T=10^3\,\mathrm{yr}$.
This aligns with theoretical estimates, as the damping of the warp can be approximated with linear theory ${\tau_\mathrm{liner\ theory}=1/(\alpha\Omega_\mathrm{out})\approx2\times 10^4\,\mathrm{yr}}$, where $\Omega_\mathrm{out}$ is the Keplerian frequency evaluated at the outer edge of the disk \citep{Lubow2000}.
In addition to viscosity, a hydrodynamic instability called `parametric instability' \citep{Gammie2000} is known to enhance the dissipation of the warp amplitude \citep{Deng2021, Fairbairn2023, Held2025}. Our analysis of the vertical velocity in our simulations suggests that the parametric instability may be present at early times before the close encounter (see Appendix~\ref{sec:parametric-insta}). At later times, however, the perturbation in the vertical velocity induced by the fly-by becomes so strong that it would mask any such instability, even if it were present. Additionally, the resolution in our simulation is too low to resolve the parametric instability in detail. Thus, we are unable to conclusively determine whether the instability is present.

The period of the warp can be estimated with the wave speed of the warp ${\tau_\mathrm{warp\ wave}=\Delta r / v_\mathrm{warp}\approx700\,\mathrm{yr}}$, where ${\Delta r = r_\mathrm{out} - r_\mathrm{in}}$ and ${v_\mathrm{warp}=c_\mathrm{s}/2}$\ \citep[e.g.,][]{Nixon2016}.
Specifically, the fact that the warp period found in our simulation is longer means that the warp wave is propagating slower than the propagation speed predicted by analytical theory. This difference partly arises as the analytic prediction was derived under the assumption of a completely inviscid disk \citep{PapaloizouTerquem1995, Ogilvie2006}. Further non-linear effects might contribute to the difference.

We notice a period change after about ${1.6 \times 10^4\,\mathrm{yr}}$ and stop the fit of the warp decay at that time.
A period change is not physically expected in the evolution of an undriven warp \citep[see e.g.][]{Kimmig2024}.
In this case of a fly-by event, the change could be caused by a change in disk structure. However, it could also be a numerical effect.

In summary, inclined fly-bys can excite warps and twists that remain even after the perturber has reached a distance ${>1000\,\mathrm{au}}$. This can mean for observations that for warps triggered by a fly-by, the fly-by object is not necessarily associated with the perturbed system.
We find moderate warp strengths of ${\leq 4^\circ}$. However, the warp can be stronger for different fly-by parameters, i.e. a shorter distance of closest approach, a more massive fly-by object, or more inclined trajectories.
Although it would be interesting to find parameters to maximize the excited warp, it is not the focus of this study.

\subsection{Spirals} \label{sec:spirals}
\begin{figure}[ht!]
    \centering
    \includegraphics[width=\linewidth]{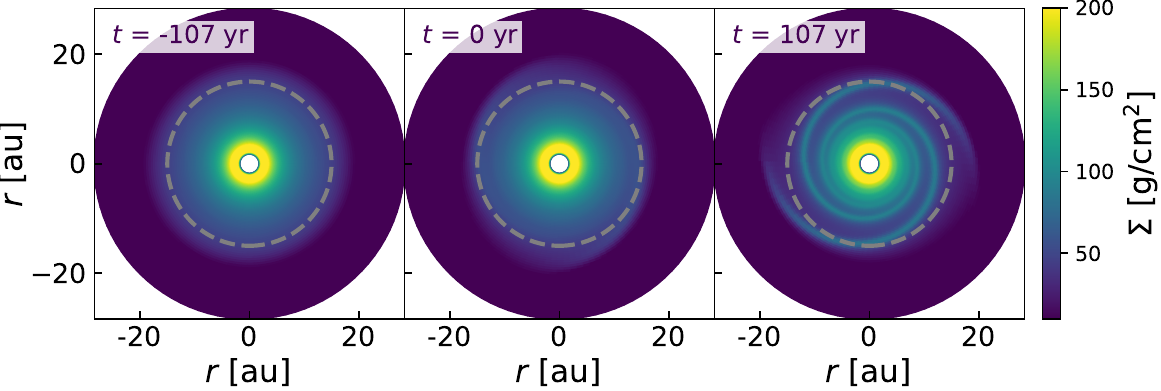}
    \caption{Surface density snapshots of the prograde fly-by in Configuration~2. Time ${t=0\,\mathrm{yr}}$ is the point of closest approach. The grey dashed circle indicates a radius of $15\,\mathrm{au}$, where we perform further analysis of the spirals. Further snapshots are shown in Appendix~\ref{sec:appendix-spirals}.}
    \label{fig:spiral-snapshots}
\end{figure}

Fly-bys are known to create spirals due to tidal effects \citep{Clarke1993, Ostriker1994}.
Figure~\ref{fig:spiral-snapshots} shows snapshots of the vertically integrated density (per radial shell) before, during, and after the close encounter of the prograde simulation in Configuration~2.
In addition to spirals, fly-bys are also known to truncate the disk. However, the disk in our simulation is already compact in the beginning, which is why we do not find truncation (see Appendix~\ref{sec:appendix-truncation}).

\begin{figure}[ht!]
    \centering
    \includegraphics[width=\linewidth]{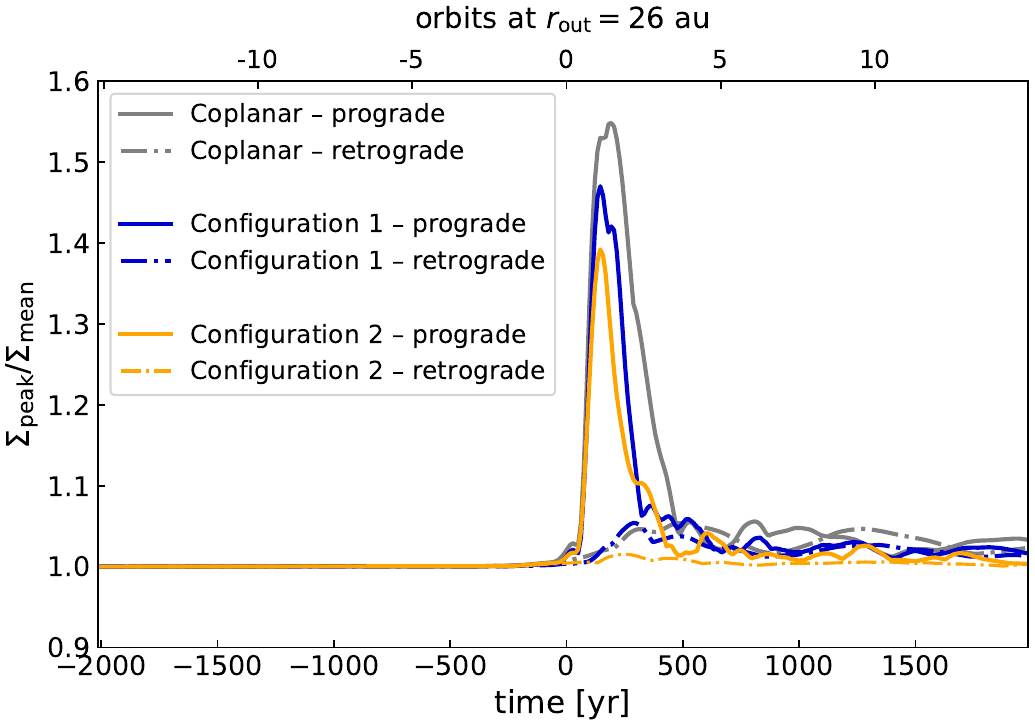}
    \caption{Time evolution of the peak surface density at $r=15\,\mathrm{au}$ in all six simulations. Prograde simulations are shown with solid lines, while retrograde simulations are indicated with dash-dotted lines.}
    \label{fig:spiral-strength}
\end{figure}

We investigate the spirals in more detail by measuring the strength of the spirals in all six different setups.
For that, we extract the maximum value of the surface density $\Sigma_\mathrm{peak}$ at a single radial location over time, analogous to \citet{Smallwood2023}, and scale it with the mean surface density $\Sigma_\mathrm{mean}$ at that radius.
Figure~\ref{fig:spiral-strength} shows the evolution of this peak surface density, which indicates the strength of the spirals, in all of our simulations at ${r=15\,\mathrm{au}}$.

We find that strong spirals occur for all three prograde setups. The retrograde simulations, on the other hand, show little or no spirals, consistent with previous findings \citep[e.g.][]{Clarke1993, Cuello2019, Cuello2023}.
We thus focus on the prograde simulations for the further evaluation in this section.

The strongest spirals occur for the coplanar prograde orbit. The simulation with the orbit tilted with respect to the $x$-axis (Configuration~1), where the periastron is in-plane, follows with the second strongest spirals.
Configuration~2, with a periastron outside of the disk plane, shows the weakest of the three prograde fly-bys.
This is consistent with the findings of \citet{Smallwood2023}, see their Figure~14 (but we note that their setup is defined rotated by $90^\circ$, so that their rotation about the $x$-axis corresponds to the case with the periastron out-of plane).

\begin{figure}[ht!]
    \centering
    \includegraphics[width=\linewidth]{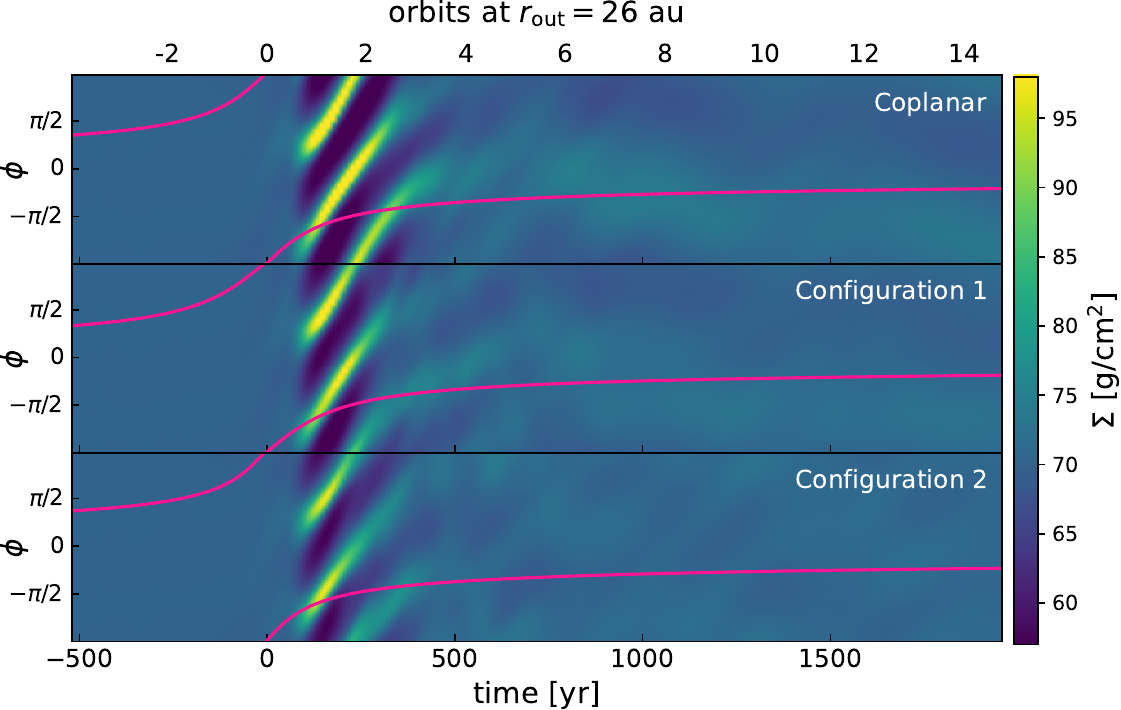}
    \caption{Azimuthal cut of the vertically integrated surface density at $r=15\,\mathrm{au}$ for all three prograde simulations. The pink line indicates the angle to the perturber projected to the $x$-$y$-plane. The time axis is scaled such that $t=0\,\mathrm{yr}$ indicates the point of closest approach.}
    \label{fig:spiral-azimuthal}
\end{figure}

We then investigate the azimuthal evolution of the spirals in Figure~\ref{fig:spiral-azimuthal}, where we plot an azimuthal cut at the same radius, $r=15\,\mathrm{au}$, over time.
In all simulations, the spirals are extremely short-lived and only last about $500\,\mathrm{yr}$.
At the time where the spirals dissolve, the perturber is located at a distance of about $370\,\mathrm{au}$ from the disk-hosting star, which are roughly $3.6\,r_\mathrm{p}$.

In comparison to \citet{Smallwood2023}, the spirals in our simulations dissolve faster on physical timescales.
As the disk in their simulations has properties different from our setup, we perform a quick analysis of timescale estimations here.
\citet{Smallwood2023} find a lifetime of about $5000\,\mathrm{yr}$ for a disk ranging from $10-100\,\mathrm{au}$. The outer disk has an orbital time scale of about $1000\,\mathrm{yr}$, which means that the outer disk performs about 5 orbits during the lifetime of the spirals.
In our simulations, the spirals live roughly $500\,\mathrm{yr}$ in a disk ranging from $2-26\,\mathrm{au}$, leading to an orbital period of $130\,\mathrm{yr}$ at the outer disk edge, which are slightly less than 4 orbits during the spiral lifetime. This compares well with the 5 orbits found by \citet{Smallwood2023}.
In this context, we want to point out that the fact that the spirals live longer than the longest orbital timescale in the disk indicates that the spiral shape is not purely caused by the Keplerian shear of a density wave.

Additional differences between our simulations and \citet{Smallwood2023} are the disk viscosity, which is smaller by at least a factor of 2 in our simulations
and the flaring of the disk. \citet{Smallwood2023} set up an unflared (often called flat) disk with a constant aspect ratio of ${H_\mathrm{p}/r=0.05}$, whereas we set our temperature structure such that the disk is flared with an index of $0.25$.
These factors could also have an impact on the lifetime of the spirals.
Furthermore, \citet{Smallwood2023} focus on less massive perturbers than we do, but for the lifetime of the spirals they do not find any strong dependency on the perturber mass.

The lifetime of the spirals in our simulations coincides with the lifetime of the additional features in the inclination profile for the prograde simulations (as seen in Figure~\ref{fig:flyby-incl}).
If the spirals are warped, they could cast shadows.
As the spirals significantly change on observable timescales, this could possibly lead to an evolution of shadow features on short timescales, as observed in a few disks, for example in TW~Hya \citep{Debes2017}, HD~135344B \citep{Stolker2017}, or MWC~758 \citep{Ren2020}.

All in all, we find that the spirals are much more short-lived than the warp. This means for observations, that the fly-by scenario can not ruled out as origin for observed warps, even in the absence of a nearby companion or an accompanying spiral pattern.

\section{Observational signatures: Models of RW~Aur} \label{sec:rwaur}

In this part of this work, we aspire to create links between our numerical models and observations.
Observationally, ongoing stellar fly-bys are proposed for quite a few systems, for example in SR~24 \citep{Weber2023, Mayama2010, Mayama2020, Fernandez-Lopez2017}, FU~Ori \citep{Takami2018, Perez2020, Borchert2022}, Z~CMa \citep{Dong2022}, UX~Tau \citep{Menard2020, Zapata2020}, AS~205 \citep{Kurtovic2018}, and more
\citep[see Tab.~1 and Fig.~6 in][]{Cuello2023}.
Because a general investigation of observational signatures of fly-bys requires a large parameter space of disk orientations with respect to the observer (which would go beyond the scope of this work), we decided to apply our models to a specific observed system: RW~Aur \citep{Ghez1993}.

The RW~Aur system is a two-star system, where both stars host a disk \citep{Cabrit2006, Rodriguez2018, Long2019} with a projected separation between the stars of about $233\,\mathrm{au}$.
A recent study by \citet[][hereafter Kur24]{Kurtovic2024} performed a detailed orbital fitting using astrometric data from different epochs, which offers good constraints for our models.
Kur24 find an eccentricity close to ${e\approx1}$, which could either be a highly eccentric bound orbit or a close-to-parabolic unbound orbit.
For the sake of this work, we will assume an unbound trajectory, even though Kur24 find that bound orbits are slightly favored.

Kur24 present ALMA dust continuum observations with high angular resolution of the two disks around A and B. For the geometry of the disks, they perform MCMC fitting and find disk inclinations of $i_\mathrm{A}=55^\circ$ and $i_\mathrm{B}=64^\circ$. Both disks have a position angle of about $\mathrm{PA}_\mathrm{B}=39.5^\circ$, and under the assumption that the near side is the same for both disks, the mutual inclination can be determined from the difference of inclinations to be about $9^\circ$.

In their analysis of the observations, Kur24 find indications of a warp in the disk around RW~Aur~A, as they find a peculiar pattern in the residuals between the observations and a planar (unwarped) parametric model.
Kur24 are able to minimize the residuals with a model including a misaligned inner disk\footnote{Kur24 fit both the inner and outer disk freely. However, the outer disk inclination and position angle remain almost the same as for their original MCMC fitting.} of $3\,\mathrm{au}$ with an inclination of $i_\mathrm{A, inn}~=~60.8^\circ$ and a position angle of $\mathrm{PA}_\mathrm{A,inn}~=~35.5^\circ$. This results in a misalignment between inner and outer disk of roughly $6^\circ$.
A warp in RW~Aur~A was even suspected prior to this study \citep{Bozhinova2016, Facchini2016rwaur}, as observations suggested a mismatch between the inclination of the inner and outer disks (inner disk\footnote{The inner disk radius here is determined from near-IR observations and is closer in than the $3\,\mathrm{au}$ by Kur24.}: $77^\circ$ by \citealt{Eisner2007}, $>60^\circ$ by \citealt{McJunkin2013}; outer disk: $45^\circ-60^\circ$ by \citealt{Woitas2001} and \citealt{Rodriguez2013}).

Millimeter observations of CO gas in the system additionally reveal a large-scale structure connecting the two stars \citep{Cabrit2006} and a chaotic environment \citep[][Kur24]{Rodriguez2018}.
\citet{Cabrit2006} proposed this large-scale structure to be a tidal arm caused by a recent close encounter of star~B with the disk around star~A, which has later been supported by hydrodynamical simulations \citep{Dai2015}.

\subsection{Hydrodynamic models} \label{sec:rw-aur-models}

In this section, we present hydrodynamic models as described in Section~\ref{sec:numerical-method}.
For the disk around RW~Aur~A, we set up an initially planar disk with a range from $1.2 - 20\,\mathrm{au}$ and an inner grid domain extended to $1\,\mathrm{au}$.
In order to keep the resolution consistent with the simulations in Section~\ref{sec:parameter-study}, we increase the number of radial cells to 160.
The orbital period at the outer edge of the disk is $80\,\mathrm{yr}$.
The chosen disk size corresponds to the size of the observed dust disk at ${\lambda=1.3\,\mathrm{mm}}$ in the system (Kur24), as we aim for radiative transfer models in the millimeter wavelength range tracing the dust continuum.
We note however, that the size of the observed gas disk is larger by a factor of $\sim 2$.
Additionally, observations suggest that the disk extends to an inner edge at ${0.1\,\mathrm{au}}$, while our inner radius is set to ${1.2\,\mathrm{au}}$.
However, smaller cavities become computationally costly in grid-based simulations, which is why we settled for a larger inner radius.

For the masses of the stars, we use ${M_\mathrm{A}=1.238\,M_\odot}$ and ${M_\mathrm{B}=0.995\,M_\odot}$ of star~A and B, respectively, leading to a mass ratio of ${q=M_\mathrm{B}/M_\mathrm{A}=0.8037}$.
For the gas disk mass, we assume ${M_\mathrm{gas}=0.01\,M_\mathrm{A}}$. We keep the slope of the surface density and the temperature structure (and hence the vertical shape of the disk) the same as in our previous models.
This gives a surface density of ${\Sigma_0=179\,\mathrm{g/cm^2}}$ at the reference radius of ${r_0 = 5.2\,\mathrm{au}}$.
As in the previous models, we set the disk viscosity to ${\alpha=10^{-3}}$.

For the geometric setup of the fly-by, we use the parameters for disk and orbit reported by Kur24 with inclination ${i_\mathrm{d,\ out} = 54.8\deg}$ and position angle of ${\mathrm{PA_{d,\ out}}=39.4\degree}$ for the outer edge of the disk and an orbit inclination ${i_\mathrm{o}=129.8\degree}$, a position angle ${\Omega_\mathrm{o}=73.8\degree}$, and an argument of periapsis ${\omega_\mathrm{o}=42.3\degree}$. The distance of closest approach is ${r_\mathrm{p}=55\,\mathrm{au}}$.
We note that these orbital parameters are given in the reference frame of the sky.
For the grid-based hydrodynamical simulations, however, it is best for the disk to initially lie in the $x$-$y$-plane.
We therefore need to compute the respective geometry of the orbital parameters with respect to the disk plane\footnote{Due to the definition of the angles as shown in Figure~\ref{fig:flyby-geometry}, we need to take the inclination ${\theta_\mathrm{d,\ calc} = 180^\circ - \theta_\mathrm{d,\ measured}}$ to calculate the mutual angles.}.
The detailed steps for the calculation of the angles can be found in Appendix~\ref{sec:mutual-geometry}.
For the eccentricity, Kur24 report $e{=0.787}$ for their best fit. In our models, we set $e=1$ for an unbound orbit.

We find a mutual inclination between the orbit and the outer disk plane to be ${\theta_\mathrm{mut}=-27.5^\circ}$, where the minus sign indicates that the periastron is below the disk plane in our simulations. This value for the mutual inclination is also reported by Kur24.
For the argument of periastron with respect to the disk plane, we find ${\omega_\mathrm{mut}=86.3^\circ}$.
At this point, we can set the longitude of ascending node arbitrarily, as the disk is initially axisymmetric.
However, this angle will become important for the radiative transfer simulations, where we need to project the model to the sky plane in order to calculate synthetic observations.
Because we do not expect a mean disk tilt of more than a few degrees, we set up the models such that the disk initially lies in the plane of the observed outer disk.

\subsubsection{Warping of the disk}

\begin{figure}[ht!]
    \centering
    \includegraphics[width=\linewidth]{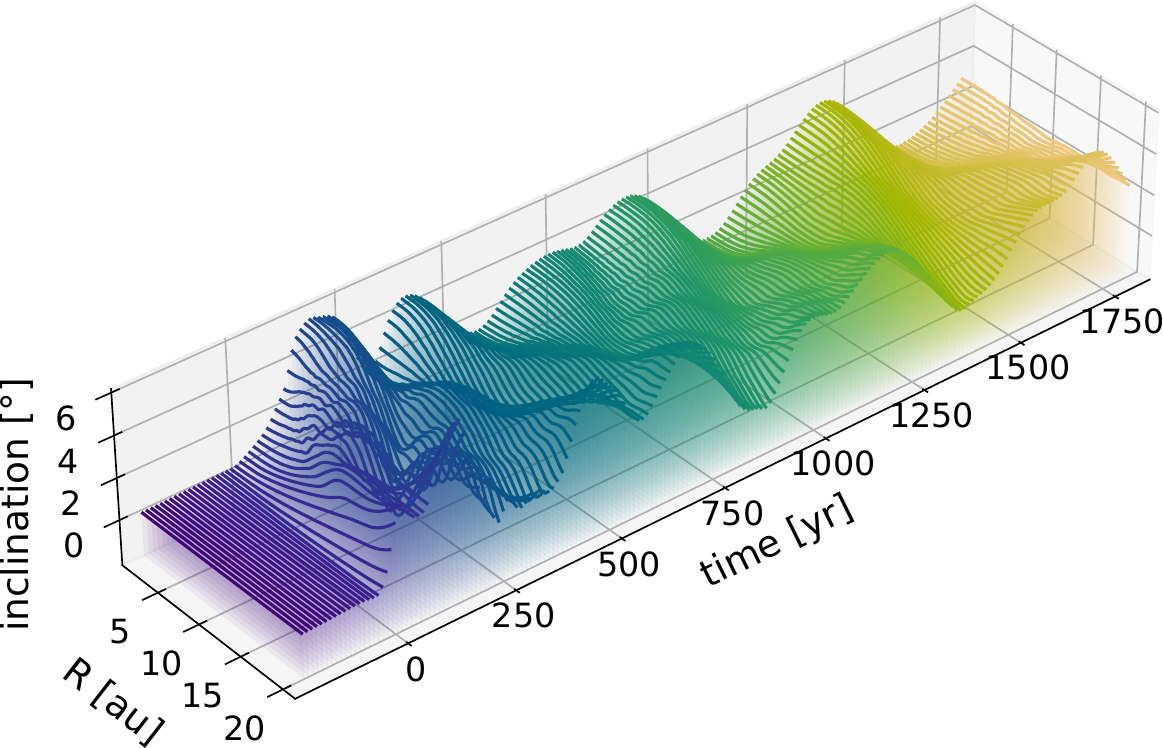}
    \caption{Evolution of the inclination profile for the simulation of the disk around RW~Aur~A. Color indicates the time, time $t=0$ indicates the point of closest approach.}
    \label{fig:rwaur-inclevol}
\end{figure}

Figure~\ref{fig:rwaur-inclevol} shows the evolution of the inclination profile for the simulation adapted to RW~Aur.
The disk develops a warp with a maximum misalignment of about $5^\circ$, which aligns nicely with the observed indications of a $6^\circ$ misalignment between inner and outer disk found by Kur24. As expected, the warp travels as a wave through the disk in our simulations.
We find a final mean inclination (defined as the angle between the total angular momentum vector and the $z$-axis) of $2.6^\circ$.
The evolution of the inclination profile shows a pattern that suggests that multiple superimposed bending waves are present.

\begin{figure}[ht!]
    \centering
    \includegraphics[width=\linewidth]{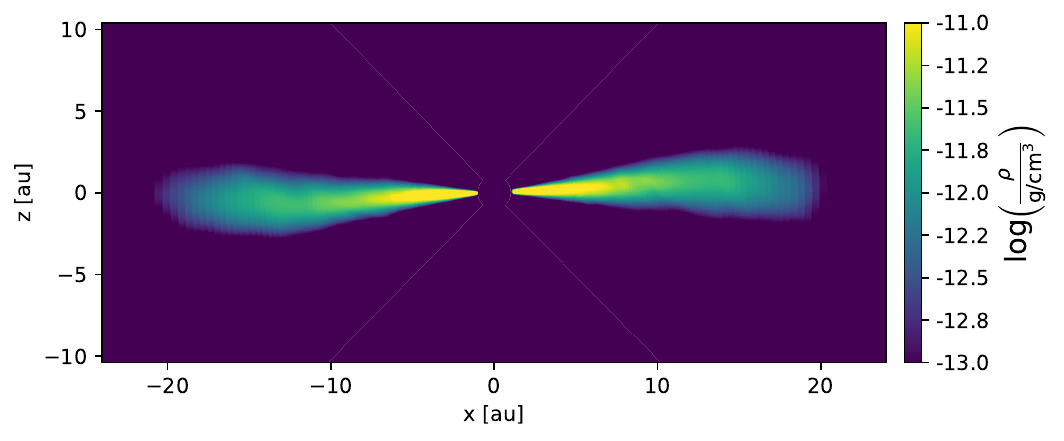}
    \caption{Cross-section of the gas density in the model of RW Aur A at the current time (295\,yr after periastron).}
    \label{fig:rwaur-crosssec}
\end{figure}

Compared to Figure~\ref{fig:flyby-incl}, the warp is slightly stronger and the wave evolves on shorter timescales.
Although the geometry of the fly-by trajectory is similar to Configuration~2 in the previous part of this work, there are some important differences.
First, the disk is smaller in size, with an outer radius of $r_\mathrm{out}=20\,\mathrm{au}$ instead of $26\,\mathrm{au}$.
This for example influences the warp timescale, which is sensitive to the disk size because of the reflection of the warp wave at the disk edges.
Further, the host star has a larger mass, whereas the perturber has a lower mass than in the previous simulations.
The distance of closest approach is shorter with $r_\mathrm{p} = 55\,\mathrm{au}$ instead of $104\,\mathrm{au}$.
Figure~\ref{fig:rwaur-crosssec} shows a cross-section through the disk at the current point of time, where the warped structure is visible.

\subsubsection{Spirals}

\begin{figure}[ht!]
    \centering
    \includegraphics[width=\linewidth]{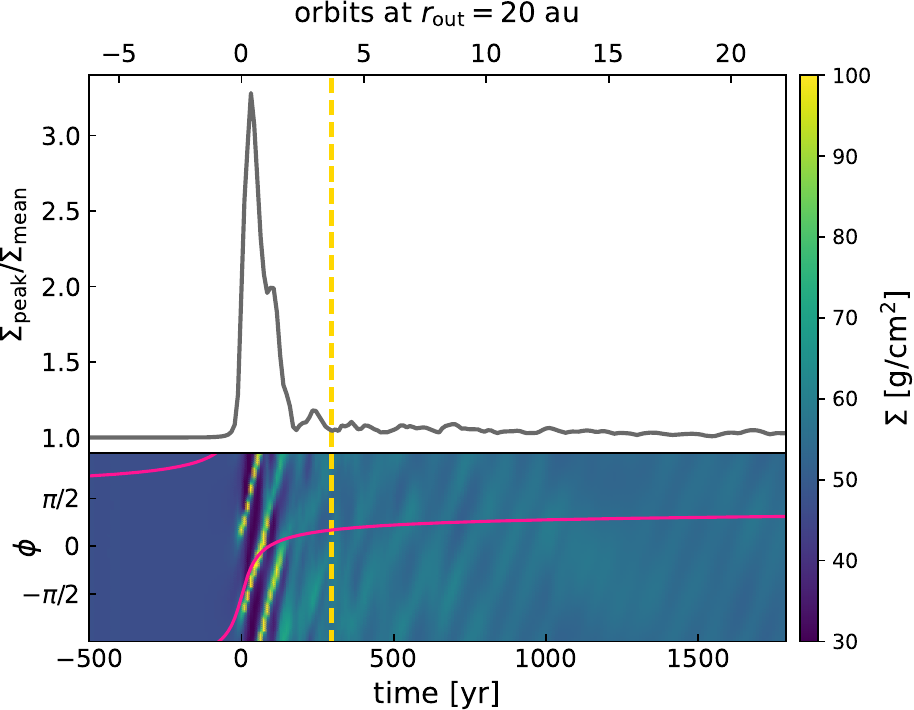}
    \caption{Peak surface density (top panel) and azimuthal profile evolution (bottom panel) of the RW~Aur simulation at $r=15\,\mathrm{au}$. As in Figure~\ref{fig:spiral-azimuthal}, the pink line indicates the angle to the perturber. The yellow dashed line highlights the current point of time at $t=295\,\mathrm{yr}$ after periastron.}
    \label{fig:rwaur-spirals}
\end{figure}

As the fly-by in the RW~Aur system is prograde, we expect the excitation of spiral arms.
Figure~\ref{fig:rwaur-spirals} shows the peak surface density (top), as well as the azimuthal density evolution (bottom) in the disk at a radius of $r=15\,\mathrm{au}$.

We see that clear spirals are excited roughly at the point of closest approach. As before, the spirals are short-lived. They disappear after roughly $200\,\mathrm{yr}$, which corresponds to 2.5 orbits at the outer disk edge. At this time, the perturber is about $200\,\mathrm{au}$ away from the disk-hosting star.
This is slightly shorter than the four orbits of the outer disk edge we found in Figure~\ref{fig:spiral-azimuthal}.
The likely reason for this are the aforementioned differences in the model setup of disk size, disk and stellar masses, and trajectory parameters.
We keep in mind that the absolute strength of the spirals in the top panel is not trivially comparable to the previous simulations, as we chose the same plotting radius of $r=15\,\mathrm{au}$ for both parts, but the disk size differs.

\subsubsection{Comparison to other work}

Hydrodynamic models of a fly-by in the RW~Aur system were performed a decade ago by \citet{Dai2015} using SPH.
Their focus was the observed large-scale tidal arm that seems to link star~A and B.
One of their main results was that a prograde encounter was necessary in order to produce such a tidal arm, which aligns very well with the findings of the astrometric orbit fitting by Kur24.
In order to produce the tidal arm, they set up an initially larger disk of $60\,\mathrm{au}$, which is then truncated by the fly-by.
When comparing the orbital parameters, their best model is similar to the result of the astrometric fit, and therefore to the setup in our work, as shown in Table~\ref{tab:orbitalparams}.
The good match of the orbital parameters is not a given, as \citet{Dai2015} fitted the orbital parameters so that their model matched the observed large-scale tidal arm.

\begin{table}[ht!]
    \centering
    \caption{Comparison of simulation setups}
    \begin{tabular}{l|p{2.25cm}|l}
        &   this work\newline \citep[and][]{Kurtovic2024}  & \citet{Dai2015} \\ \hline
    \textbf{fly-by parameters} & & \\
    stellar mass $M_\mathrm{A}$ & $1.238\,M_\odot$ & $1.4\,M_\odot$ \\
    stellar mass $M_\mathrm{B}$ & $0.995\,M_\odot$ & $0.9\,M_\odot$ \\
    eccentricity $e$ & 1 & 1 \\
    closest approach $r_\mathrm{p}$  & $55\,\mathrm{au}$ & $70\mathrm{au}$ \\
    mutual inclination $\theta$ & $27^\circ$  & $18^\circ$ \\
    arg. of periastron $\omega$ & $89^\circ$ & $80^\circ$  \\ 
    & & \\
    \textbf{disk parameters} & & \\
    initial disk size & $20\,\mathrm{au}$ & $60\,\mathrm{au}$ \\
    viscosity $\alpha$ & $10^{-3}$ & $5 \times10^{-3}\ ^{(a)}$ \\ 
    initial disk mass & $1.238 \times 10^{-2}\,M_\odot$ & $1.6 \times 10^{-3}\,M_\odot$ \\
    \end{tabular}
    \tablefoot{(a) The actual viscosity might be higher, since estimations in SPH of physical viscosity from artificial viscosity often underestimate this parameter \citep{Meru2012}.}
    \label{tab:orbitalparams}
\end{table}

Comparing the hydrodynamic results of the disk in detail is difficult, as the disk properties differ strongly, also shown in Table~\ref{tab:orbitalparams}.
A similarity we find are the excited spirals, which is not surprising given the prograde nature of both fly-by setups.
Looking at their Figure~4, the spirals also seem to be short-lived and seem to dissolve within $200\,\mathrm{yr}$ after periastron passage.
A direct comparison of the warping is not possible, as they did not explicitly analyze the warp.
However, we expect the warp evolution to differ for models of different viscosity \citep{Kimmig2024}. In the simulations by \citet{Dai2015}, the viscosity is larger by at least a factor of 2.

However, the fact that the trajectory parameters are similar might indicate that we could reproduce a similar large-scale tidal structure in our simulations if our initial disk size (and the computational domain) was larger.

\subsection{Radiative transfer models of the dust continuum} \label{sec:dust-continuum}

In this section, we aim to compare our models to the observations.
For this, we run Monte Carlo simulations using the radiative transfer code RADMC-3D \citep{radmc3d}.
The dust temperature is calculated consistently with the radiative transfer model using the Monte Carlo \citet{BjorkmanWood2001} method implemented in {RADMC-3D}, where we use the modified random walk option \citep{Fleck1984}, that accelerates the computation of photon packages in optically very thick regions by applying the analytical solution of the diffusion equation. We use $10^8$ photon packages to compute the temperature.
We then compute the sky model of the system with ray-tracing (also $10^8$ photon packages) and produce synthetic images by convolving the fluxes with a two-dimensional Gaussian imitating the corresponding telescope's finite resolution.
We use a beam size of $18\times30\,\mathrm{mas}$ and a position angle of $8.186\degree$, as reported for the observational data.
We note that to remain consistent with the observations by Kur24, we use a value of $154\,\mathrm{pc}$ for the distance to RW~Aur~A \citep{Gaia2016, Gaia2021}, instead of the $156.1\,\mathrm{pc}$ found in the Gaia Data Release 3 \citep{Gaia2023}.

We compare the simulation with the ALMA dust continuum observations.
For the composition of the dust, we assume the DIANA standard composition of 87\% carbon and 13\% pyroxene, where the pyroxene is composed of molecule bonds with 70\% magnesium and 30\% iron.
The porosity is set to 25\% \citep{Woitke2016}.
We compute the dust opacity and scattering matrices with \texttt{optool} \citep{optool}.

Since the hydrodynamic simulation only models gas, we need to assume a spatial dust distribution in the disk.
Because the behavior of larger dust particles in warping disks is not well known yet, we assume small dust grains of a singular size $a=1\,\mu\mathrm{m}$ that are perfectly coupled to the gas.
The choice of a singular dust size in contrast to a dust size distribution is motivated by the lack of good observational constraints on the actual size distribution. Additionally, it allows us to ensure that the dust follows the dynamics of the gas.

For the dust-to-gas ratio, we aim to match the total flux of the observation, which Kur24 reported to be ${34.5\pm 0.01\,\mathrm{mJy}}$.
With a dust-to-gas ratio of $10^{-2}$ (and a distance of $154\,\mathrm{pc}$ to the source), our radiative transfer model results in a total flux of $31.5\,\mathrm{mJy}$, and we therefore adopt this dust-to-gas ratio.
However, we note that our choice of a single grain size of small dust could underestimate the flux.
We also note that the dust-to-gas ratio is somewhat flexible, as the total gas disk mass in our hydrodynamic simulations is scalable, as we are not treating self-gravity, nor the indirect term from the disk onto the star.
This means that the
same total flux can also be reached when the gas disk mass and dust-to-gas ratio are scaled differently.

\begin{figure}[ht!]
    \centering
    \includegraphics[width=0.9\linewidth]{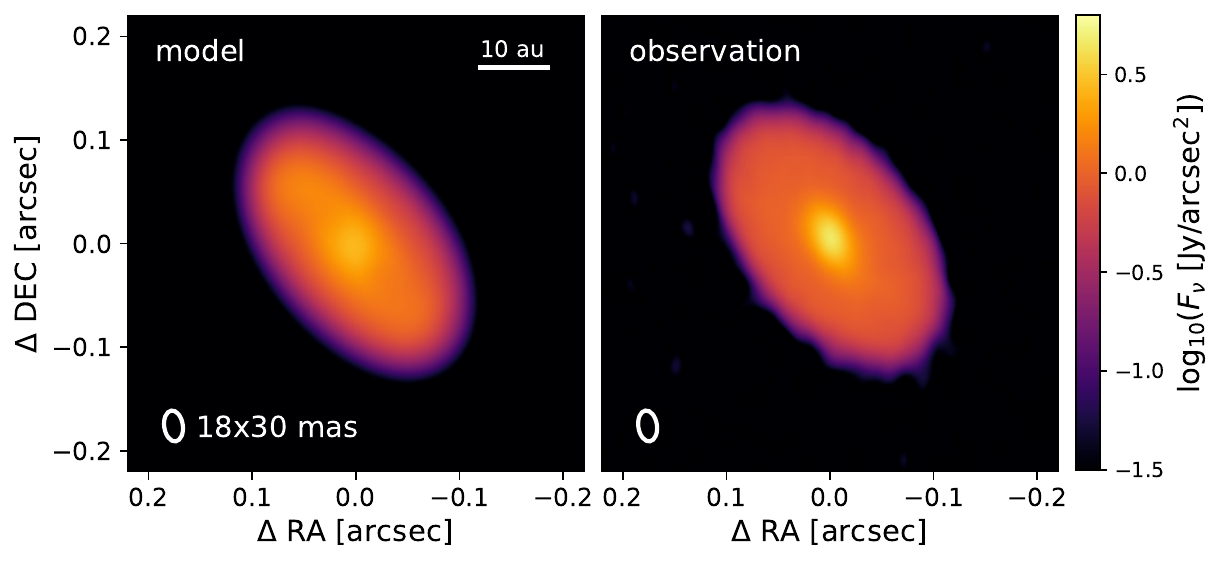}
    \caption{Synthetic observation of our hydrodynamic simulation at ${t=295\,\mathrm{yr}}$ after periastron at a wavelength of $\lambda=1.3\,\mathrm{mm}$ (left panel) next to the ALMA Band~6 observation of RW~Aur \citep[right panel,][]{Kurtovic2024}, also at $\lambda=1.3\,\mathrm{mm}$. The ellipse in the bottom left corner indicates the beam and is the same for both images.}
    \label{fig:rwaur-observation}
\end{figure}

According to the fit by Kur24, the time of observation was ${t=295\,\mathrm{yr}}$ after the periastron. We confirmed in our simulation that the position of the perturber matches the observed position of RW~Aur~B.
Figure~\ref{fig:rwaur-observation} shows a comparison of the model next to the observation by Kur24. Both images are observed at a wavelength of ${\lambda=1.3\,\mathrm{mm}}$.
Overall, the images compare well.
In the observations, the emission in the central beam is stronger roughly by a factor of 1.5, which could be because of our choice of a larger cavity in the simulation for reasons of computation time. 
We present the comparison between the convolved and unconvolved model in Appendix~\ref{sec:unconvolved}, which shows that a few spiral features close to the inner edge are visible, which are then hidden by the beam resolution.
A fine-tuned quantitative match is out of the scope of this work, as many parameters, such as disk characteristics, fly-by trajectory and dust models, have an impact on the result of the model.

\begin{figure}[ht!]
    \centering
    \includegraphics[width=0.9\linewidth]{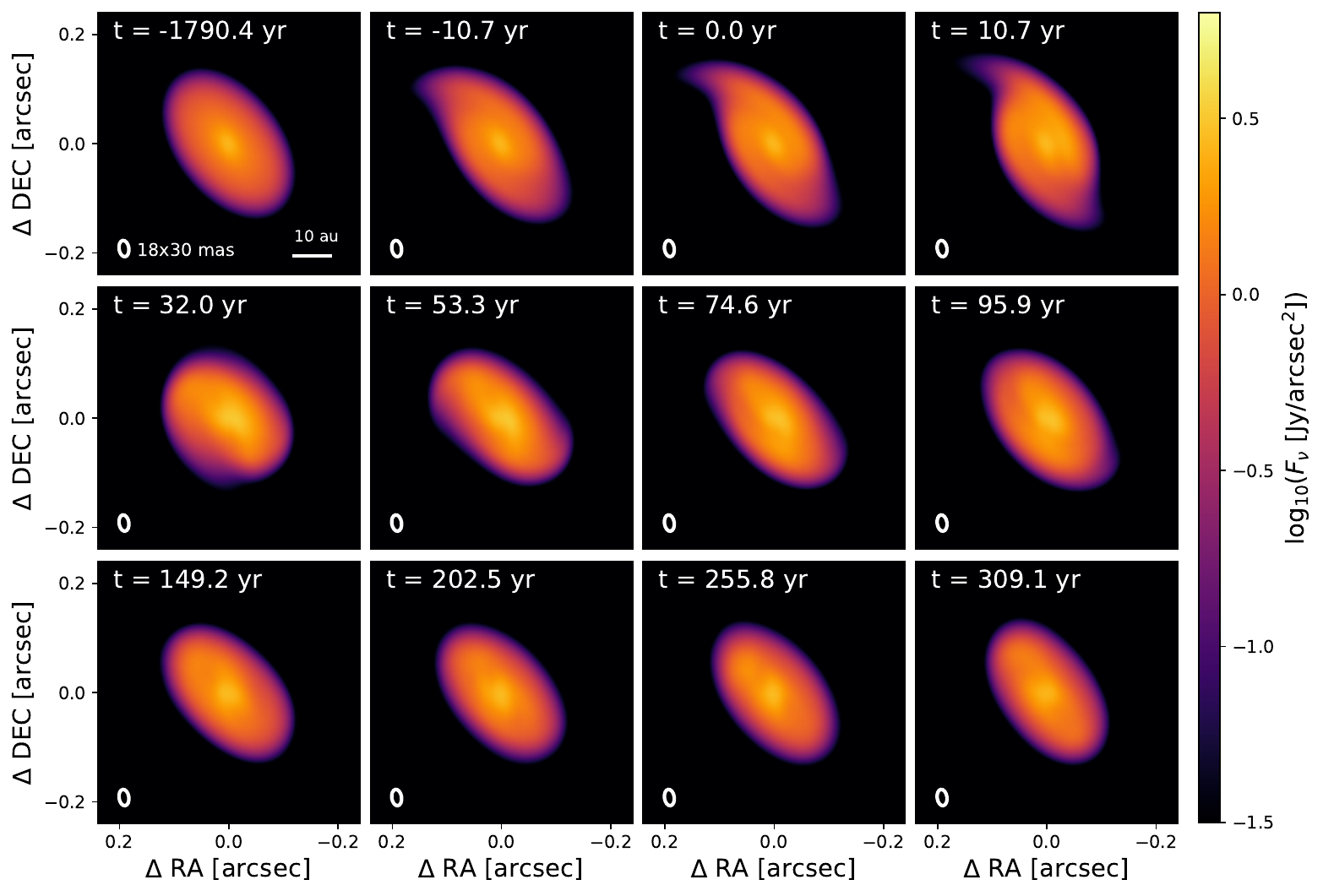}
    \caption{Synthetic observations at $\lambda=1.3\,\mathrm{mm}$ of the RW~Aur~A simulation at different times. The first panel shows the initial setup of the simulation, time $t=0\,\mathrm{yr}$ is the time of the closest approach, and the current time step lies between the second last and last panels. The images are convolved with the same beam as Figure~\ref{fig:rwaur-observation}.}
    \label{fig:rwaur-radtrans-timeevol}
\end{figure}

In Figure~\ref{fig:rwaur-radtrans-timeevol}, we present radiative transfer simulations at different times of the hydrodynamic simulation.
During the point of closest approach (${t=0\,\mathrm{yr}}$), the spiral structure excited by the fly-by is indeed visible, but quickly disappears.
The system is (synthetically) observed from the same location at each time, which corresponds to inclination and position angle of the initial setup of the simulation.
The warp only changes the disk plane by a few degrees throughout the simulation, and therefore the observed inclination of the disk does not vary strongly.
However, a slight variation of the inclination and position angle on a time scale of a few tens of years could be observable.

One of the objectives of the comparison between the simulations and the real observation is whether the spiral structures produced by the fly-by should be observable with state-of-the-art instruments.
At $t=295\,\mathrm{yr}$ in our simulations, the strongest spirals have already dissolved, as we show in Figure~\ref{fig:rwaur-spirals}. In the simulations by \citet{Dai2015}, the spirals seem to be weak as well at the time of comparison.
However, some weak spiral structures are still residue in our simulations, also seen in the unconvolved image in Figure~\ref{fig:unconvolved-radtrans}.
After the convolution (Figure~\ref{fig:rwaur-observation}), however, we find that these structures are not discernible, leading to the good match between model and observation.

\subsection{Synthetic gas kinematics}

Observations of molecular lines can give insight into the gas dynamics of protoplanetary disks \citep[see e.g.][]{Pinte2023}.
The kinematic structure can be highly complex \citep{Teague2025}.
We investigate the kinematic structures in the RW Aur system due to the recent close encounter in line radiative transfer models using RADMC-3D.

We choose the $J{=}2{-}1$ rotational transition of $^{12}\mathrm{CO}$ at ${\lambda=1.3\,\mathrm{mm}}$ and computed 200 channels in a window of ${\pm 20\,\mathrm{km/s}}$ around the molecular line.
For the molecular abundance, we assumed a number density fraction ${^{12}\mathrm{CO}/\mathrm{H}_2=10^{-4}}$.
We took into account the CO-depletion via freeze-out for temperatures lower than $20\,\mathrm{K}$ \citep{Williams2014}. For that, we assumed the gas to be thermally coupled with the dust and computed the dust temperature using the Bjorkmann \& Wood method.
We further included the photodissociation effect for column number densities lower than a threshold value \citep{Visser2009}.
We chose a threshold of ${N_\mathrm{dissoc}=5 \times 10^{19}\,\mathrm{H}_2/\mathrm{cm}^2}$, which is in alignment with the range found by \citet{Trapman2023} and \citet{Rosotti2025}.
To compute the vertical column density, we used an interpolation from our spherical grid. Then, we set the abundance to zero wherever the column density drops below the threshold.
To save computation time, we neglect scattering effects for the line radiative transfer.

\begin{figure}[ht!]
    \centering
    \includegraphics[width=0.9\linewidth]{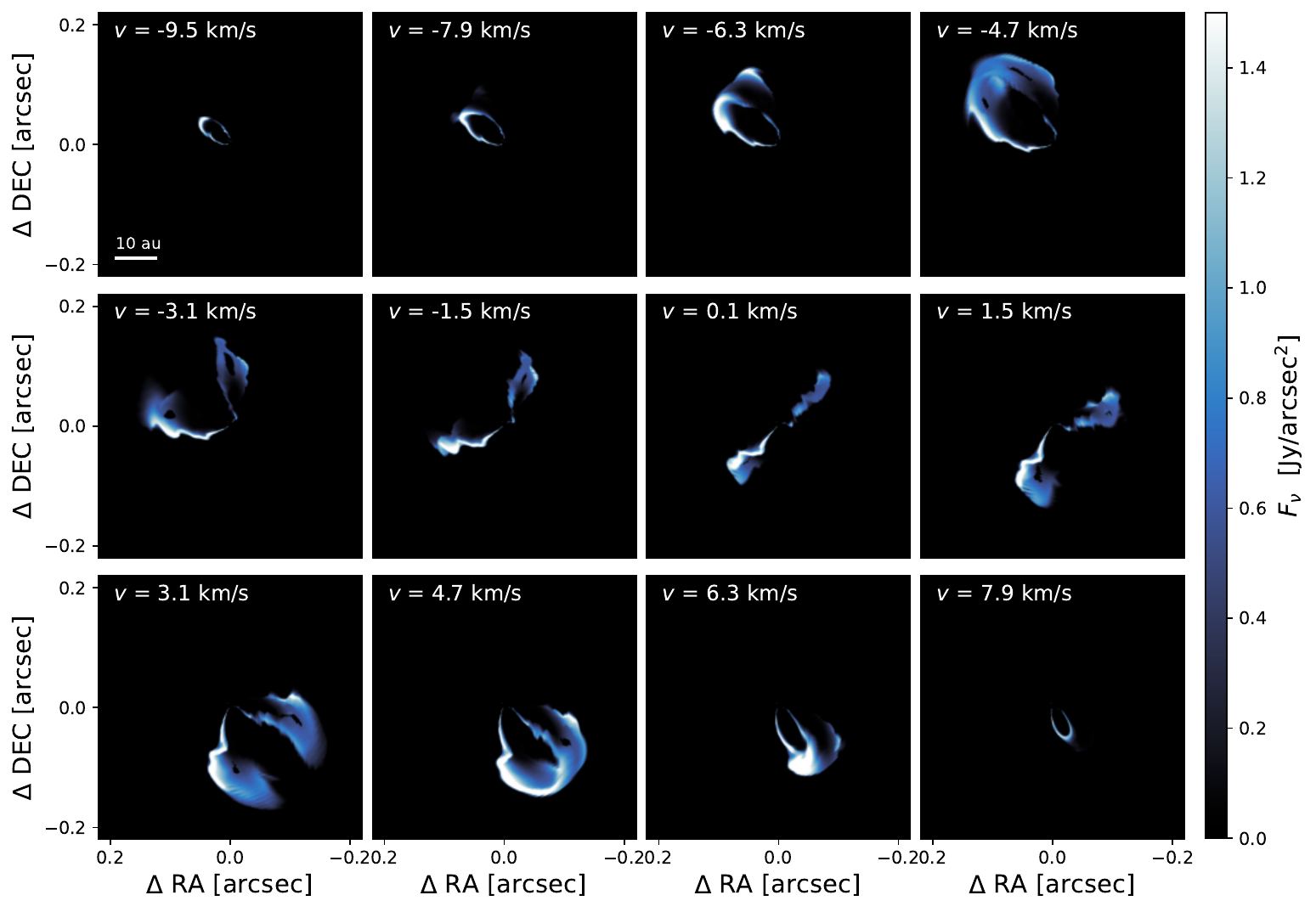}
    \caption{Intensity of different velocity channels of the RW~Aur~A model at the current time of $295\,\mathrm{yr}$ after periastron passage.}
    \label{fig:rwaur-butterfly}
\end{figure}

Unfortunately, a detailed comparison to actual observations of the RW~Aur system is not useful at this point, as the currently available kinematic observations do not possess sufficient spectral resolution to show details in the kinematics (see e.g. Kur24).
We keep the kinematic images unconvolved for now, and all moment maps are created with the unclipped data. In this section, we focus on the current time at $295\,\mathrm{yr}$ after periastron passage.

Figure~\ref{fig:rwaur-butterfly} shows some of the resulting velocity channels from the line radiative transfer model.
Recall that the upper right side (north-west) is the near side of the disk.
Many channels show interesting `wiggly' features that clearly deviate from a usual model of a smooth disk.
We investigate these features further by computing moment maps.

\begin{figure}[ht!]
    \centering
    \includegraphics[width=0.8\linewidth]{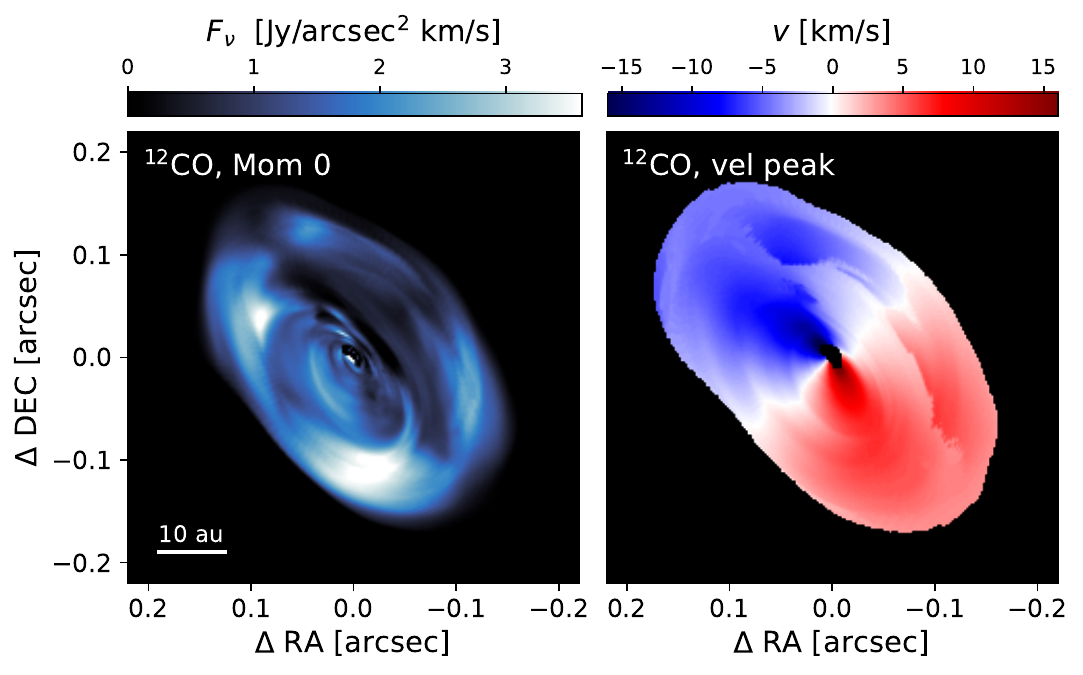}
    \caption{Moment~0 map (left) and velocity of the peak intensity (Moment~9, right) of the gas in the hydrodynamic model of the disk around RW~Aur~A at $295\,\mathrm{yr}$ after periastron. The line is centered at ${v=0\,\mathrm{km/s}}$, which corresponds to proper motion corrected observations.}
    \label{fig:rwaur-momentmaps}
\end{figure}

\begin{figure}[ht!]
    \centering
    \includegraphics[width=0.6\linewidth]{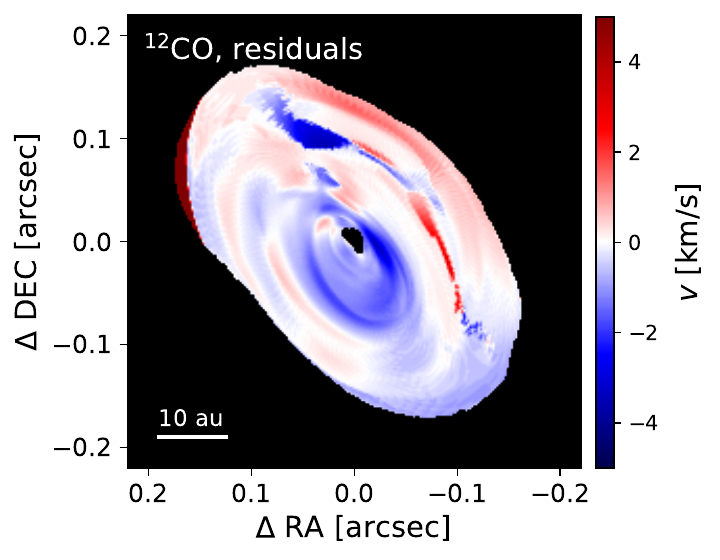}
    \caption{Deviation from a Keplerian profile of the Moment~9 map shown in the right panel of Figure~\ref{fig:rwaur-momentmaps}.}
    \label{fig:rwaur-kinematic-residuals}
\end{figure}

The left panel of Figure~\ref{fig:rwaur-momentmaps} shows the Moment~0 map.
In contrast to the dust continuum images presented in Section~\ref{sec:dust-continuum}, it clearly shows a disturbed structure.
The emission of the observed $^{12}\mathrm{CO}$-line originates in the upper layers of the disk, as opposed to the dust continuum, which typically traces the disk midplane.
This indicates that the disk surfaces are more disturbed than the disk midplane.
In the map of the velocity of the peak intensity (also called Moment~9 map, Figure~\ref{fig:rwaur-momentmaps}, right), the ${v=0\,\mathrm{km/s}}$-line (white) also appears to be wiggly.
To investigate this in more detail, we subtracted a Keplerian disk model to reveal the residuals, shown in Figure~\ref{fig:rwaur-kinematic-residuals}.
For the Keplerian disk model, we used the initial setup of our hydrodynamic simulation, for which we computed a kinematic data cube in the same way using RADMC-3D. The velocity of the peak intensity of this is shown in Appendix~\ref{sec:unconvolved}, Figure~\ref{fig:rwaur-keplerian-model}.

The deviation from a Keplerian model is significant and
we find complex dynamical features in the residuals.
Some of these features might originate from remnants of the spirals that were excited during the fly-by event.
The warped morphology of the disk may also affect the kinematic structure.
While a further detailed analysis to disentangle these effects is highly relevant in the current development of the field \citep[e.g.][]{Pinte2023, Teague2025}, it would go beyond the scope of this work, and we thus have to leave it for future work.
At this point, we also want to note that the observed gas disk of RW~Aur~A is larger than in our hydrodynamic model. We chose the more compact size in order to match the dust continuum observations when assuming that the dust is evenly distributed.
For a better analysis of the kinematics in the RW~Aur system with predictions for observations, considering a larger gas disk could be helpful.

However, the results of this section show that better kinematic observations of the RW Aur system could very well reveal remnant features of the recent close encounter and provide further insight into the dynamics of the system.

\section{Discussion} \label{sec:discussion}

\subsection{Fly-bys as potential origin of moderate warps}

Our hydrodynamical models show that warps of a few degrees are a natural consequence of fly-bys with an inclined trajectory with respect to the disk plane.
The persistence of the warp beyond the actual fly-by encounter raises the fly-by scenario as possibility for the origin of observed moderate warps, even if no perturber can be dynamically associated with the system. 
In fact, moderate warps might be the rule rather than the exception in protoplanetary disks. For example, \citet{Winter2025} found indications of warps in the kinematics of almost all disks of the exoALMA sample \citep{Teague2025}.
In alignment with that, \citet{Villenave2024} observed asymmetries in near IR observations in 15/20 disks observed edge-on. These asymmetries are likely created by moderate warps of a few degrees \citep{Kimmig2025}.

That said, the damping of the warp highly depends on the disk viscosity \citep[][see also Appendix~\ref{sec:appendix-rwaur-viscosities}]{Nixon2016, Kimmig2024}. This means that the warp persists longer than the fly-by event only in low-viscosity disks ($\alpha \leq 10^{-3}$ as rough reference). Therefore, it is crucial to account for low-viscosity scenarios in models, which highlights the need for grid-based simulations to complement SPH models, that usually cannot represent physical viscosities of $\alpha$ below a few times $10^{-3}$ \citep[see][]{Lodato2010, Meru2012}.

However, an inclined stellar fly-by is not the only possible cause for a warped disk and the origin of moderate warps remains unknown.
Alternative formation scenarios include misaligned infall of material onto the disk \citep{Thies2011, Dullemond2019, Kuffmeier2024}, misaligned magnetic fields with respect to the disk plane \citep{Foucart2011, Romanova2021}, radiation-induced warping (for stars with $L \gtrsim 10\,L_\odot$) \citep{Pringle1996, Armitage1997}, misaligned binaries \citep{Facchini2013, Lodato2013, Deng2022, Rabago2024}, or outer bound companions such as stars or planets \citep{Papaloizou1995, Nealon2018, Zanazzi2018, Zhu2019}.

\subsection{Implications and caveats in the RW~Aur case}

For the application to RW~Aur, we assume an unbound fly-by trajectory, even though \citet{Kurtovic2024} find that bound orbits are more likely by a factor of 0.28.
In general, the effect of highly eccentric orbits is expected to be similar to that of unbound orbits \citep{Cuello2023}.
We can perform a quick timescale estimation: The best fit of \citet{Kurtovic2024} includes an eccentricity of $e=0.78$ and leads to an orbital period of approximately $2800\,\mathrm{yr}$.
In our simulations, the excited warp is expected to be still present after this time, which means that a repeated encounter due to the bound nature of the orbit could create a different warp structure than a single fly-by.
However, since the maximum amplitude of the warp is only of a few degrees and strongly dampened during one orbital period, our results might still be viable even for the bound case.

In the event of a close encounter, strong UV radiation from the fly-by perturber can influence the dynamics of the disk \citep{Guarcello2023}.
This UV radiation radiation can cause photoevaporation \citep{Winter2022}, which plays a decisive role in disk evolution \citep[e.g.,][]{Clarke2007, Facchini2016photoevap, Winter2018, Keyte2025}. However, in the case of RW~Aur~A, we do not expect strong radiation onto the disk, as the perturber is of relatively low mass. Nevertheless,
shadows in the disk due to the warp could influence the dynamics by changing the temperature structure of the disk \citep{Su2024, Zhang2024, Ziampras2024}.

Interestingly, a jet is observed in the RW~Aur system \citep{Hirth1994, Alencar2005, Melnikov2009}.
Jets are thought to be linked to magnetic fields and could indicate MHD winds launched from the disk. Magnetic fields likely influence the dynamics of warped disks. So far, warping of protoplanetary disks under the influence of magnetic fields has not been well studied.
Jets can also replenish the interstellar material in these environments, which can affect nearby disks that accrete this material for example through streamers \citep{Pineda2020, Ginski2021, Codella2024, Cacciapuoti2024}.

An interesting fact worth mentioning is that RW~Aur~A is found to have an unusually high accretion rate of about ${10^{-7}\,M_\odot/\mathrm{yr}}$ \citep{Hartigan1995}.
\citet{Cabrit2006} suggested the recent fly-by as possible explanation, as fly-bys could be able to trigger episodes of increased accretion \citep[so-called FU~Orionis events][]{Bonnell1992, Pfalzner2008, Forgan2010}.
Even though strictly speaking RW~Aur~A is not an FU~Orionis object, according to the formal definition \citep[see e.g.][]{Connelley2018}, accretion episodes could still play a role for the star and disk.
Additionally, several dimming events have been observed for RW~Aur~A \citep{Rodriguez2013, Rodriguez2016, Antipin2015, Petrov2015} which have been suggested to be caused by surrounding disrupted material \citep{Rodriguez2013, Dodin2019} or a warp in the disk \citep{Bozhinova2016, Facchini2016rwaur}.

\section{Conclusion} \label{sec:conclusion}

In this work, we performed simulations of inclined fly-bys passing a protoplanetary disk, where we put the focus on the warping of the disk.
In our simulations, we chose to model parabolic orbits ($e=1$), where the velocity at the periastron is lowest in comparison to other unbound orbits. This means that parabolic fly-bys are expected to have the largest influence on the disk.

\begin{itemize}
\item[•] We find that inclined fly-bys can trigger a warp of a few degrees, which evolves in a wave-like manner through the disk, lasting much longer than the fly-by itself. The warping is most prominent in fly-bys where the periastron is not in the same plane as the disk, especially in a retrograde fly-by.
This is particularly interesting, as the warping in such fly-by configurations has not been studied well before.

\item[•] The lifetime of spirals, that are excited especially in prograde configurations, is much shorter than that of the warp, with the spirals lasting only about four orbits of the outer disk.
Observationally, this means that a warp triggered by a fly-by could still be observed when the fly-by object is long out of sight and no obvious spiral structures are observed in the disk. In other words, fly-bys may provide an explanation for some of the observed moderate warps.

\item[•] We applied the stellar fly-by scenario to the RW~Aur system, where a recent (about $300\,\mathrm{yr}$ ago) close encounter between two stars hosting a disk is suspected.
Recent orbital fitting by \citet{Kurtovic2024} gives good constraints on the mutual geometry of the fly-by trajectory with respect to the disk, which we can use to set up our simulations.
We find that the trajectory of RW~Aur~B around the RW~Aur~A is likely to induce a warp of around $5^\circ$ in the disk around star~A, consistent with observational indications of a misalignment between inner and outer disk regions of $6^\circ$.

\item[•] Radiative transfer simulations of our hydrodynamic model enable a comparison of the resulting dust continuum to the observations.
We find that in dust continuum wavelengths ($\lambda=1.3\,\mathrm{mm}$) the disk in our model looks smooth and the spirals are not visible, which compares well to the observation of the dust continuum of RW~Aur~A.
We further find interesting features in the synthetic CO-line images that hint toward valuable insights that can be gained from future kinematic observations of this system.
\end{itemize}

In summary, this work shows that fly-bys might be able to explain frequently observed warps and shadow features in protoplanetary disks. Depending on disk properties, the warp can remain present for longer timescales, so that the perturber might not always be observable.

\begin{acknowledgements}
We kindly thank N. Kurtovic, T. Rometsch, L.-A. Hühn, I. Rabago, M. Leemker, G. Lodato, H. Klahr, and R. Klessen for their support and helpful comments.
We remember the legacy of Prof. Willy Kley, who passed away in 2021, and would have been a part of this work.
We acknowledge funding from the DFG research group FOR 2634 “Planet Formation Witnesses and Probes: Transition Disks” under grant DU 414/23-2 and KL 650/29-1, 650/29-2, 650/30-1.
GR and CK acknowledge support from the European Union (ERC Starting Grant DiscEvol, project number 101039651) and from Fondazione Cariplo, grant No. 2022-1217.
SF acknowledges financial contribution from the European Union (ERC, UNVEIL, 101076613) and from PRIN-MUR 2022YP5ACE.
Views and opinions expressed are, however, those of the author(s) only and do not necessarily reflect those of the European Union or the European Research Council. Neither the European Union nor the granting authority can be held responsible for them.
PW acknowledges support from FONDECYT grant 3220399 and ANID -- Millennium Science Initiative Program -- Center Code NCN2024\_001.
Additionally, we acknowledge support by the High Performance and Cloud Computing Group at the Zentrum für Datenverarbeitung of the University of Tübingen, the state of Baden-Württemberg through bwHPC and the German Research Foundation (DFG) through grant INST 37/935-1 FUGG.
This paper makes use of the following ALMA data: ADS/JAO.ALMA\#2018.1.00973.S ALMA is a partnership of ESO (representing its member states), NSF (USA) and NINS (Japan), together with NRC (Canada), MOST and ASIAA (Taiwan), and KASI (Republic of Korea), in cooperation with the Republic of Chile. The Joint ALMA Observatory is operated by ESO, AUI/NRAO and NAOJ.
Parts of this work were included in a PhD thesis.

\end{acknowledgements}

\bibliographystyle{aa}
\bibliography{flyby}

\begin{appendix}
\onecolumn

\section{Initial azimuthal velocity in the hydrodynamical simulations} \label{sec:azimuthalvelo}
The initial azimuthal velocity follows a Keplerian profile with a correction for pressure gradients.
Because of the exponential cut-offs in the density, we need to adjust the implementation of the azimuthal velocity in FARGO3D, as the cut-offs introduce additional pressure gradients that are not accounted for in the default implementation.
For that, we set the initial azimuthal velocity to
\begin{equation}
    v_\phi (r) = v_\mathrm{k} \left(\frac{r}{r_\mathrm{cyl}}\right)^{f} + \frac{r_\mathrm{cyl}}{\rho} \frac{\partial\ p(r_\mathrm{cyl},z)}{\partial r_\mathrm{cyl}},
\end{equation}
where $v_\mathrm{k} = \sqrt{G M_* / r}$ is the Keplerian velocity, $r$ the spherical radius and $r_\mathrm{cyl}$ the cylindrical radius in the coordinate system where the central star is at the origin, $\rho$ the density, and $p$ the pressure.
As in \citet{Dullemond2020}, Equation~C.6, we use the cylindrical radius in this pressure gradient correction.  We solve the differential quotient in the initial setup numerically with the first-order midpoint method on a fine grid.
Without this adjustment for the azimuthal velocity, the disk would spread and quickly smooth out the surface density cut-off especially at the outer disk edge, as the material would possess too much angular momentum. This effect can be seen in \citet{Kimmig2024}, Figure~B.3, where the pressure gradient due to the cut-off had not yet been accounted for.

\section{Hyperbolic trajectories} \label{sec:appendix-traj}

This section provides a reminder on the parameters to describe unbound orbits (hyperbolae) and an overview over the implementation of the initial position and velocity of the star on the fly-by trajectory, called perturber from here on, in our simulations.

\begin{figure} [ht!]
  \centerline{\includegraphics[width=0.3\textwidth]{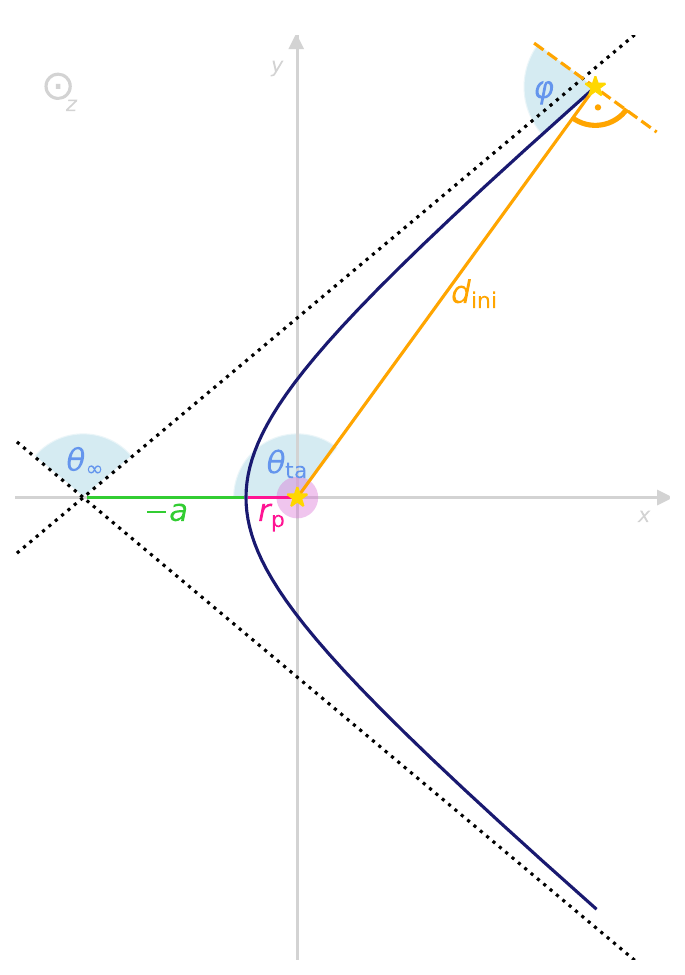}}
  \caption{\label{fig:hyperbel} Schematics of a hyperbolic trajectory (dark blue line) for a coplanar fly-by viewed from above.
  The host star with a disk (indicated by the purple circle) is placed at the origin of the coordinate system.
  Black dotted lines indicate the asymptotes.
  }
  \label{fig:flyby-schematics}
\end{figure}

The eccentricity of hyperbolic orbits is ${e\geq1}$. Orbits with ${e=1}$ are called parabolic. 
The eccentricity is related to the external angle between the asymptotes $\theta_\infty$ with ${e = - {1}/{\cos(\theta_\infty)}}$.
The distance between the periastron (also called periapsis) to the focal point is typically denoted with $r_\mathrm{p}$. In the scenario of a fly-by, this distance is often referred to as the distance of closest approach.
When eccentricity and the distance of closest approach are given, the semi-major and -minor axes can be calculated as
\begin{equation}
a = - \frac{r_\mathrm{p}}{(e-1)} \\
b = - a \sqrt{e^2-1}.
\end{equation}
Here, the semi-major axis $a$ is negatively defined, as indicated in Figure~\ref{fig:flyby-schematics}.
The hyperbolic trajectory can be expressed in the form
\begin{equation}
r = \frac{l}{1 - e \ \cos(\theta_\mathrm{ta})},
\end{equation}
where ${l=- b^2/a}$ is the semi-latus rectum and $\theta_\mathrm{ta}$ is the true anomaly, an angular parameter indicating the current position of the object along the trajectory.

FARGO3D takes the initial position and velocity vector as the fly-by object as input and integrates the trajectory using a fifth-order Runge-Kutta N-body solver.
We therefore need to calculate these quantities from the parameters of the hyperbola.
For that, we fix the initial distance $d_\mathrm{ini}$ between the perturber and the disk hosting star (indicated in Figure~\ref{fig:flyby-schematics}).

From this initial distance, we can calculate the initial true anomaly using the hyperbolic trajectory equation
\begin{equation}
\theta_\mathrm{ta, ini} = \arccos \left(- \frac{1}{e} \left[ \frac{b^2}{a\ d_\mathrm{ini}} + 1 \right] \right).
\end{equation}

For a coplanar fly-by, the plane of the trajectory coincides with the plane of the disk, which lies in the $x$-$y$-plane in our setup.
In this case, the initial position can be calculated from the true anomaly with
\begin{equation}
x  = - \cos(\theta_\mathrm{ta}) \ d_\mathrm{ini}, \\
y  = \sin(\theta_\mathrm{ta}) \ d_\mathrm{ini}, \\
z  = 0.
\end{equation}

The initial velocity vector can be determined with the flight path angle $\varphi$
\begin{equation}
\tan(\varphi) = \frac{e\ \sin(\theta_\mathrm{ta})}{1 + e\ \cos(\theta_\mathrm{ta})},
\end{equation}
leading to the velocity components
\begin{equation}
v_x = - v_\mathrm{abs} \cos \left(\varphi - \theta_\mathrm{ta} + \pi/2 \right) \\
v_y = - v_\mathrm{abs} \sin \left(\varphi - \theta_\mathrm{ta} + \pi/2 \right) \\
v_z = 0,
\end{equation}
where $v_\mathrm{abs}$ is the absolute value of the velocity. The absolute value can be calculated using the vis-viva equation
\begin{equation}
v_\mathrm{abs} = \sqrt{\mu \left( \frac{2}{d_\mathrm{ini}} - \frac{1}{a} \right)},
\end{equation}
where $\mu = G M_\mathrm{tot}$ is the standard gravitational parameter with $M_\mathrm{tot}$ as sum of the mass of the fly-by object and the mass of the host star.

Inclined fly-by trajectories can then be achieved by rotating the original in-plane hyperbola according to the orbital elements.

To test the configuration of the trajectory, we performed a couple of test simulations with different trajectory configurations. Because we were mainly interested in the fly-by trajectory in these tests, we set the disk resolution very low, which speeds up the computation time.
This is reasonable, as the N-body integrator of FARGO3D does not depend on the grid resolution for the disk.
The resulting trajectory from FARGO3D perfectly matches the analytical solution of the hyperbola in all our test cases, which confirms that our implementation of the initial position and velocity works correctly.

\section{Further evaluation of the hydrodynamic models}

\begin{figure}[ht!]
\centering
\begin{minipage}{.45\textwidth}
    \centering
    \includegraphics[width=0.9\linewidth]{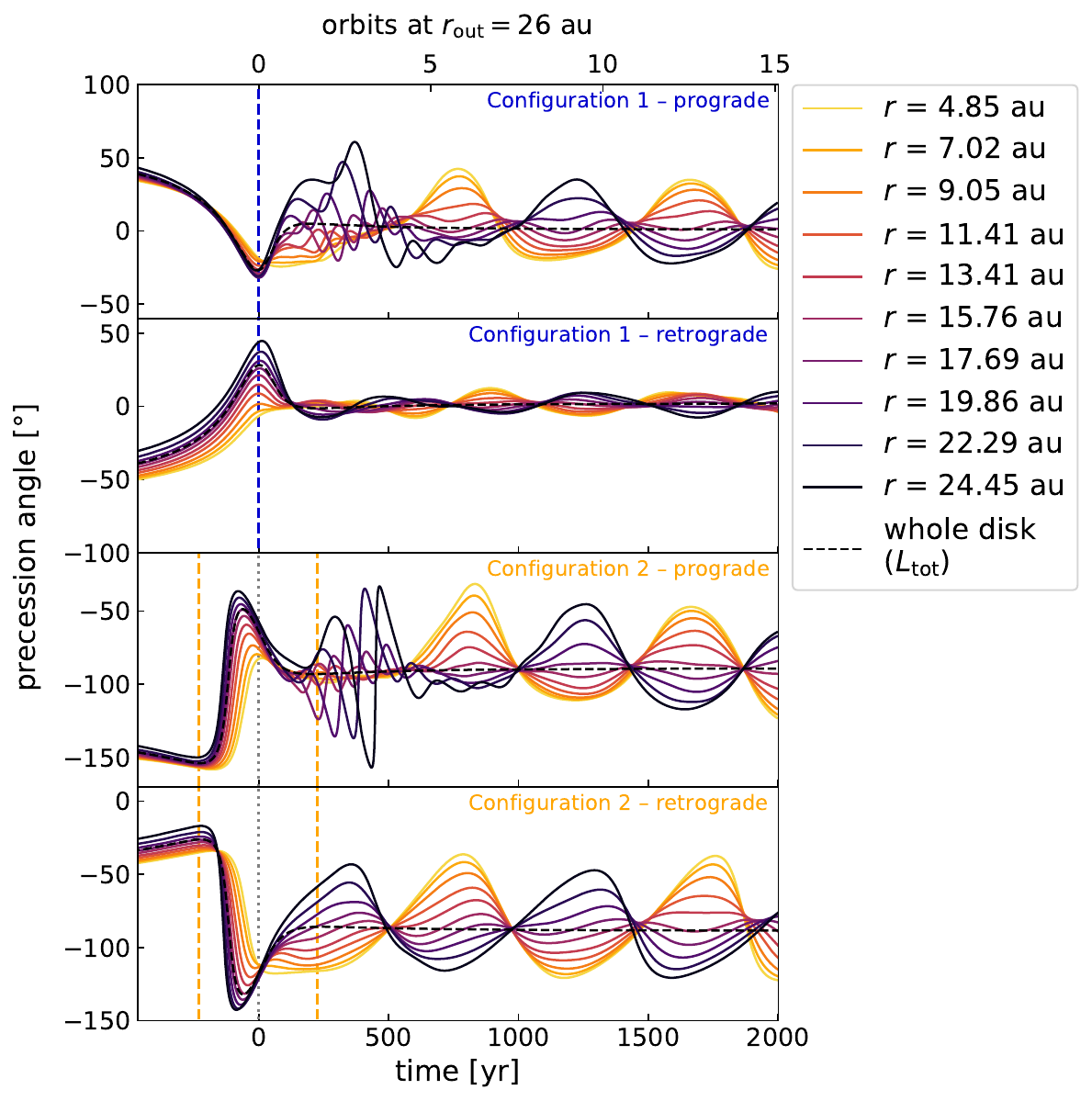}
    \caption{Precession angles for different radii in all four inclined simulations.
    The black dashed line indicates the precession angle for the total angular momentum vector of the disk. We only show times for where the inclination is $\geq 0.1^\circ$, as the precession angle is undefined otherwise. The blue and orange dashed lines highlight the times when the perturber crosses the initial disk midplane.
    For a good comparison, the range of the $y$-axis is the same in all panels.}
    \label{fig:twist}
\end{minipage} \hspace{0.5cm}
\begin{minipage}{.45\textwidth}
    \centering
    \includegraphics[width=\linewidth]{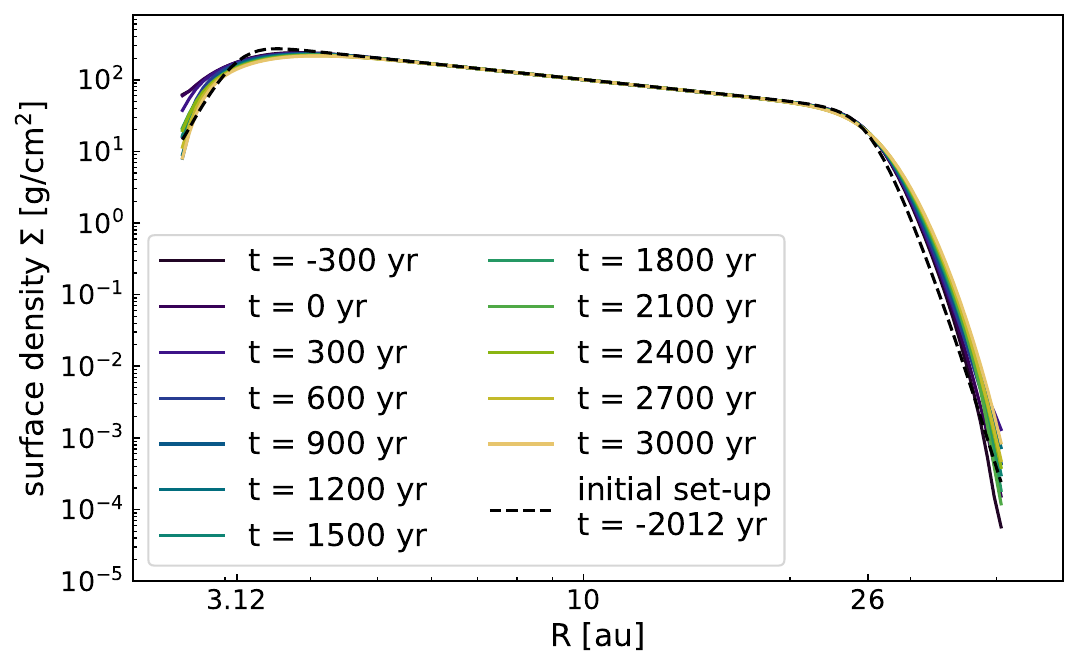}
    \caption{Evolution of the surface density in the retrograde simulation in Configuration~2. The time $t=0\,\mathrm{yr}$ corresponds to the time of closest approach. The black dashed line indicates the initial surface density setup according to Equation~\ref{eq:surfdens}.
    }
    \label{fig:surfdens}
\end{minipage}
\end{figure}

\subsection{Twist} \label{sec:appendix-twist}

In addition to a warp, the disk can twist, a differential precession described analogously to \citet{Kimmig2024} by the angle between the angular momentum vector of a disk annulus, projected to the $x$-$y$-plane, and the $x$-axis.
Twists have been found to occur in fly-by scenarios, for example in \citet{Cuello2019}.
Figure~\ref{fig:twist} shows the time evolution of the precession angle $p$, which we calculate as the angle between the $x$-axis from the projection of the angular momentum vector to the $x$-$y$-plane
\begin{equation}
    p = \mathrm{arctan}\left( L_y, L_x \right),
\end{equation}
where $L_{x,y}$ are the $x,y$-components of the angular momentum vector.

We find a differential precession, i.e., a twist, after the disk is warped, consistent with \citet{Kimmig2024}. This twist likely occurs due to pressure forces in the disk that influence the warp evolution. This is currently under investigation (Aly et al., in prep.).
We note that the differential precession of the disk occurs on a full circular motion, even though Figure~\ref{fig:twist} may suggest a motion rocking back and forth.
This is due to our definition of the precession angle with respect to the static coordinate system instead of the precession axis. However, as the precession axis changes over time, there is no unambiguous definition for the precession motion, and we therefore decided to stay consistent with the definition of \citet{Kimmig2024}.
Both prograde fly-bys show a more complicated pattern in the time between $0-500\,\mathrm{yr}$, which is likely due to the spirals.
If we compare Configuration~1 (with a periastron in plane) to Configuration~2, we see that the twist occurs slightly earlier for Configuration~2.
This likely occurs because the perturber crosses the disk plane already prior to the periastron. To highlight this, we included the blue and orange dashed lines in Figure~\ref{fig:twist}.
At these times, the gravitational force in vertical direction changes sign, leading to a strong perturbation.
The twisting is strongest for the retrograde case of Configuration~2, where the warp is also strongest.

\subsection{Disk truncation} \label{sec:appendix-truncation}
Fly-bys are known to truncate the disk. We therefore briefly discuss this phenomenon in our simulations.
\citet{Breslau2014} find an empirical relation of final disk sizes after a fly-by
\begin{equation}
r_\mathrm{final}\approx0.28\ q^{-0.32}\ r_\mathrm{peri},
\end{equation}
where $q$ is the mass ratio between the two stars and $r_\mathrm{peri}$ the distance of the closest approach. Disks experiencing a prograde fly-by are affected more strongly than retrograde fly-bys, where disks can be up to twice as large \citep{Bhandare2016, Cuello2019}.

In our simulations, $q=1$ and $r_\mathrm{peri}=104\,\mathrm{au}$ and therefore we would expect a disk size of $29\,\mathrm{au}$. However, our disks are already compact from the beginning, with $26\,\mathrm{au}$ and we therefore do not expect the fly-by to affect the size of the disk.

Figure~\ref{fig:surfdens} shows the surface density evolution for the retrograde Configuration~2 as an example. Indeed, the disk does not show truncation effects and only spreads slightly due to viscous evolution.
We find this behavior in all of our simulations.
This allows us to keep the main focus on the warp evolution of the disks discussed above.

\subsection{Spirals} \label{sec:appendix-spirals}

\begin{figure}[ht!]
    \centering
    \includegraphics[width=0.8\linewidth]{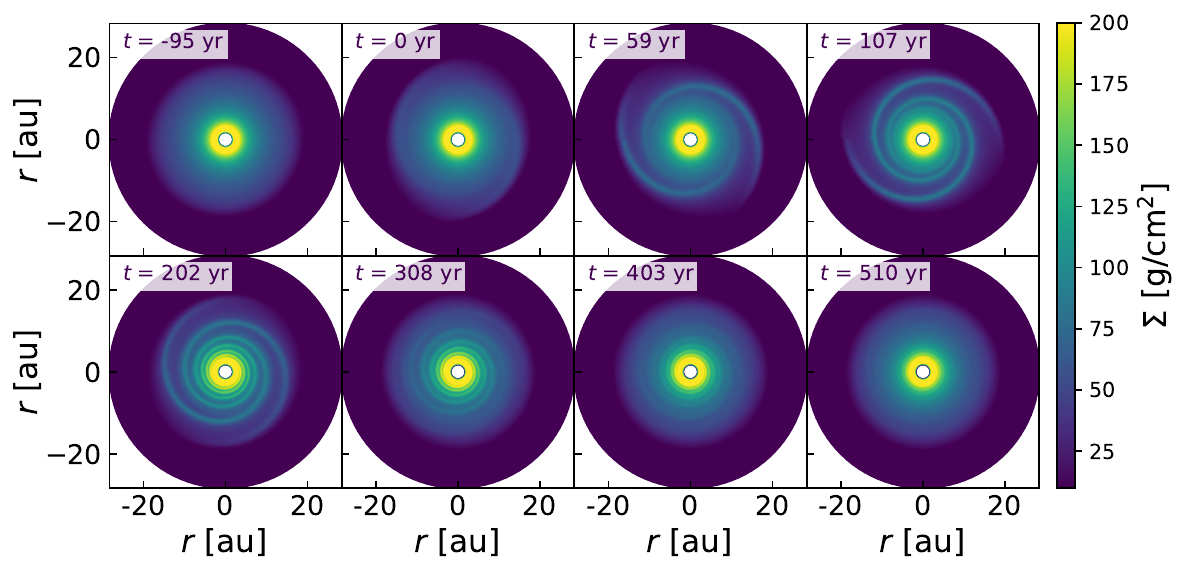}
    \caption{Snapshots of the simulation with an inclined prograde fly-by in Configuration~2 (periastron out of the disk plane). The closest approach occurs at $t=0\,\mathrm{yr}$. The color scale for the surface density is linear, which highlights the spirals better than a logarithmic scale.}
    \label{fig:spiral-different-times}
\end{figure}

Here, we present the spiral structure excited by the stellar fly-by in the prograde simulation of Configuration~2. In Figure~\ref{fig:spiral-different-times}, we show the 2D surface density, which we obtain by vertically integrating the density in each radial shell. We show different snapshots within the lifetime of the spirals of about $500\,\mathrm{yr}$, as suggested by Figure~\ref{fig:spiral-azimuthal}.
We note that for more massive disks, the lifetime of the spirals could be prolonged if self-gravity becomes important \citep{Papaloizou1989}.

In agreement with \citet{Smallwood2023}, the pitch angle decreases over time as the spirals wind up. A detailed measurement of pitch angle and pattern speed is not trivial and would go beyond the scope of this work.

\section{Parametric instability}\label{sec:parametric-insta}

Warped disks with low viscosity are suspected to be prone to the parametric instability \citep{Gammie2000}, which is a hydrodynamic instability that arises due to a parametric resonance of inertial waves and is enhanced by the free energy in oscillating shear flows. Such shear flows, often called sloshing and breathing motions, are known to occur in warped disks \citep[see e.g.,][]{Lubow2000, Ogilvie2013a, Ogilvie2013b, Fairbairn2021a, Fairbairn2021b, Dullemond2022}.
Previous studies have shown that the parametric instability can operate already for small warp amplitudes and can enhance the dissipation of the warp \citep{Deng2021, Fairbairn2023, Held2025}.
It is also known to operate in disks containing eccentricities \citep{Ogilvie2014, Dewberry2025}. This parametric instability can enhance the dissipation of the warp amplitude.

We attempt to investigate if the parametric instability is operational in our simulations. For this, we continue to focus on the retrograde simulation in Configuration~2.
We first transform the velocity field to a warped coordinate system, which aligns the midplane, meaning that the midplane is flat in these rotated coordinates. As the parametric instability is expected to be most prominent in the vertical velocity, we we focus on this velocity in a regime around the disk midplane.
We note that we subtract the value of the vertical velocity of the midplane from the data, which is likely a motion of the disk plane arising from the warp evolution. In Section~\ref{sec:appendix-breathing}, we verify that the resulting subtracted vertical velocity agrees with the expected vertical velocity from linear warp theory.

\begin{figure} [ht!]
  \centerline{\includegraphics[width=0.8\textwidth]{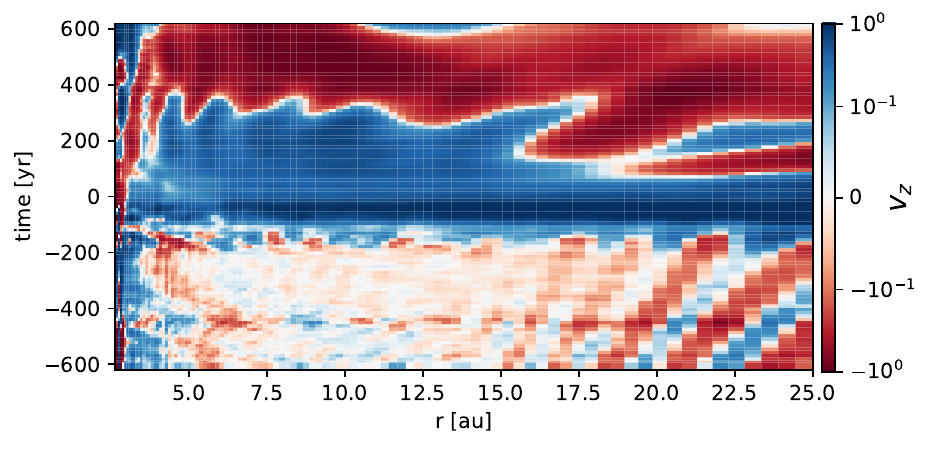}}
  \caption{\label{fig:vz-evol}
  Radial profile of the vertical velocity at $\theta=0.15$ (at a specific azimuthal location) evolving over time (vertical axis). Time $t=0\,\mathrm{yr}$ indicates the point of closest approach of the flyby. The vertical velocity is normalized by the maximum at each time step, similar to \citealt{Fairbairn2023}, Figure~10.
  We kept the rasterized grid appearance intentionally, as it highlights our resolution in space and time.}
\end{figure}

We then plot the time evolution of the radial profile of one specific altitude ($\theta = 0.15\,\mathrm{rad}$, $\phi = 0\degree$) in Figure~\ref{fig:vz-evol}. This representation of the vertical velocity is similar to \citealt{Fairbairn2023}, Figure~10, or \citealt{Dewberry2025}, e.g. Figure~4.

What is clearly visible in Figure~\ref{fig:vz-evol} is that the fly-by event adds strong distortions in the vertical directions that overpower any details that might result from the parametric instability. However, before the point of closest approach, structures appear similar to the results of \citealt{Fairbairn2023}, who investigated the parametric instability in detail on very small scales.
Motivated by this, we evaluated the warp amplitude to estimate how much the disk is warped before the periastron of the perturber. The warp amplitude is defined by
\begin{equation}
    \psi = \left| \frac{\mathrm{d}\vec{l} }{\mathrm{d\ ln} (r)} \right|,
\end{equation}
where $\vec{l}(r)$ is the local angular momentum vector.

\begin{figure}[ht!]
\centering
\begin{minipage}{.45\textwidth}
    \centering
    \includegraphics[width=\textwidth]{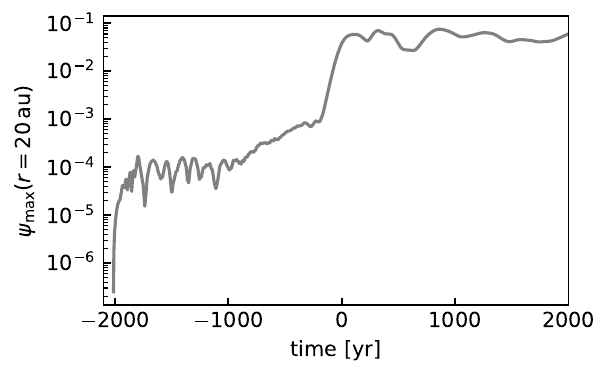}
    \caption{\label{fig:psi}Warp amplitude $\psi$ at $r=20\,\mathrm{au}$ in the disk.}
    \label{fig:twist}
\end{minipage} \hspace{0.5cm}
\begin{minipage}{.45\textwidth}
    \centering
    \includegraphics[width=0.7\textwidth]{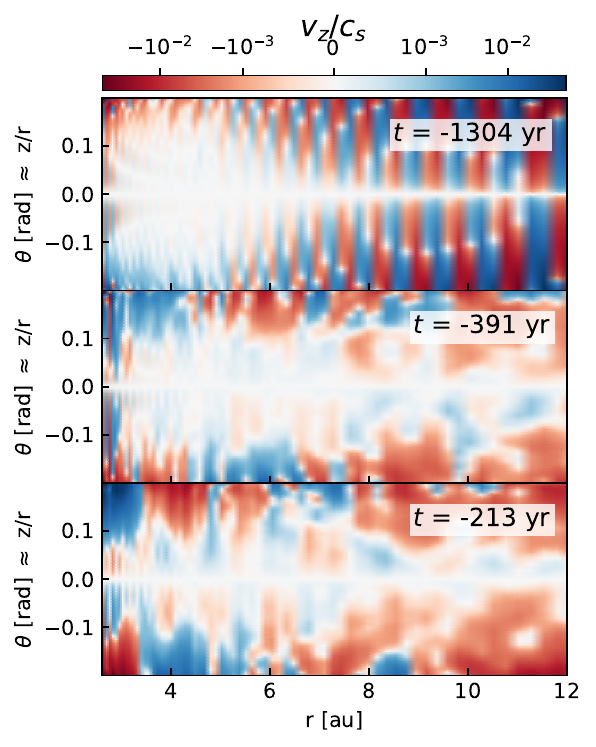}
    \caption{\label{fig:vz-snapshots}
  Cross-section of the vertical velocity $v_z$ scaled by the local sound speed $c_s$ at three different times before the moment of closest approach. The times are negative, as $t=0\,\mathrm{yr}$. Note that this is zoomed in to the inner region of the disk $r\leq12\,\mathrm{au}$.
  }
\end{minipage}
\end{figure}

Figure~\ref{fig:psi} shows that the disk already slightly warps before the moment of closest approach ($t=0\,\mathrm{yr}$) up to $\psi=10^{-3}$. This could already be enough to trigger the parametric instability \citep{Fairbairn2023}. We thus show the full vertical cross-sections of the vertical velocities at different times in Figure~\ref{fig:vz-snapshots}.

Figure~\ref{fig:vz-snapshots} shows an oscillating vertical structure that resembles the structure found in a simulation by \citealt{Deng2021} (see their Figure~6). In their work, they find indications that the parametric instability is at work in their simulations.
This structure could be a signature of the parametric instability. However, our simulations do not contain enough resolution to resolve the very fine structure found by \citealt{Fairbairn2023}, and further, more detailed analysis and simulations with a much higher resolution would be required to strongly claim the presence of the parametric instability in our simulations.

\citealt{Deng2021} find an increase in vertical kinetic energy in their simulations as an indication of the parametric instability. Evaluating the vertical kinetic energy in our simulations, we also find an increase. However, as our simulation includes an external perturber, the increase in vertical kinetic energy is connected to the perturber and therefore can not straight-forwardly interpreted as a signature of the parametric instability.

\subsection{Breathing motions}\label{sec:appendix-breathing}

\begin{figure} [ht!]
  \centerline{\includegraphics[width=0.7\textwidth]{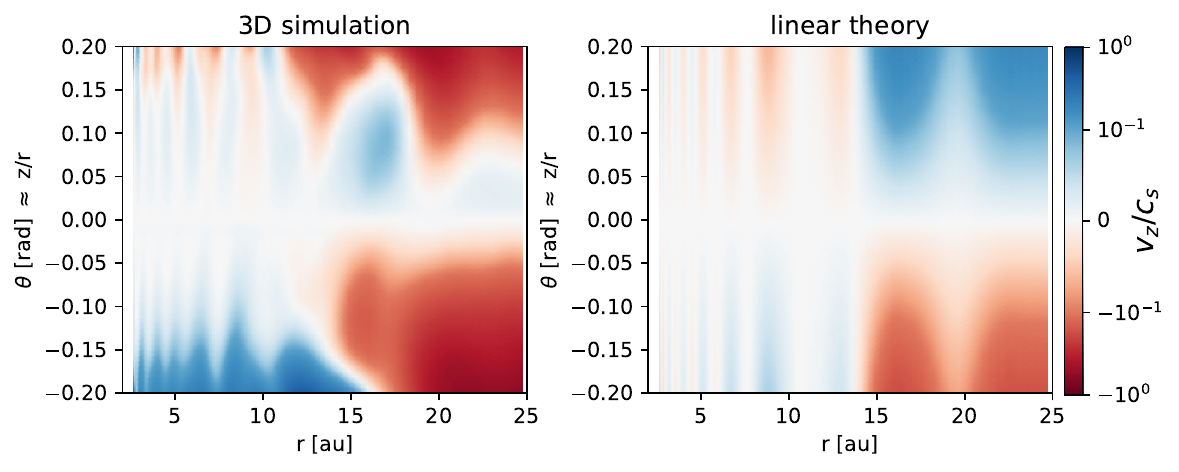}}
  \caption{\label{fig:vz-lintheo}
  Cross-section of the velocity $v_z$ scaled by the local sound speed at $t=320\,\mathrm{yr}$ after the closest approach, shown in the rotated coordinate system. Left: velocity field in the 3D simulation, where the vertical velocity of the midplane is subtracted. Right: Reconstructed breathing motions from linear theory.}
\end{figure}

In this section, we take a closer look at the breathing motions occurring in the disk \citep{Ogilvie2013a, Fairbairn2021b}. The purpose of this is mainly to verify our evaluation of the vertical velocity in Section~\ref{sec:parametric-insta}. We evaluate the retrograde simulation in Configuration~2 and aim to look at a stage when the disk is significantly warped. For that, we arbitrarily choose $t=320\,\mathrm{yr}$ after the closest approach. We present a cross-section through the vertical velocity in the rotated coordinate system in Figure~\ref{fig:vz-lintheo}, left panel.

To investigate this pattern more closely, we compute the vertical velocity component $v_z = V_z \Omega z$ according to linear analysis, given the exact warp shape at this snapshot of our simulation. Here, $V_z$ is the velocity coefficient in the linear approximation as for example in \citealt{Ogilvie2013b} and \citealt{Dullemond2022}. For details on the extraction of $V_z$ from the 3D model, we refer the reader to \citealt{Kimmig2024}, Appendix~D.
This reconstructed vertical velocity from linear theory is shown in Figure~\ref{fig:vz-lintheo}, right panel. The two panels agree well in the regions close to the midplane, where the linear theory is applicable. We find some deviations at higher altitudes, especially close to the outer edge of the disk ($r_\mathrm{out} = 26\,\mathrm{au}$), which likely result from non-linear effects in the 3D simulation.

\section{Calculation of the mutual geometry between orbit and disk} \label{sec:mutual-geometry}

\begin{figure}[ht!]
    \centering
    \includegraphics[width=0.45\linewidth]{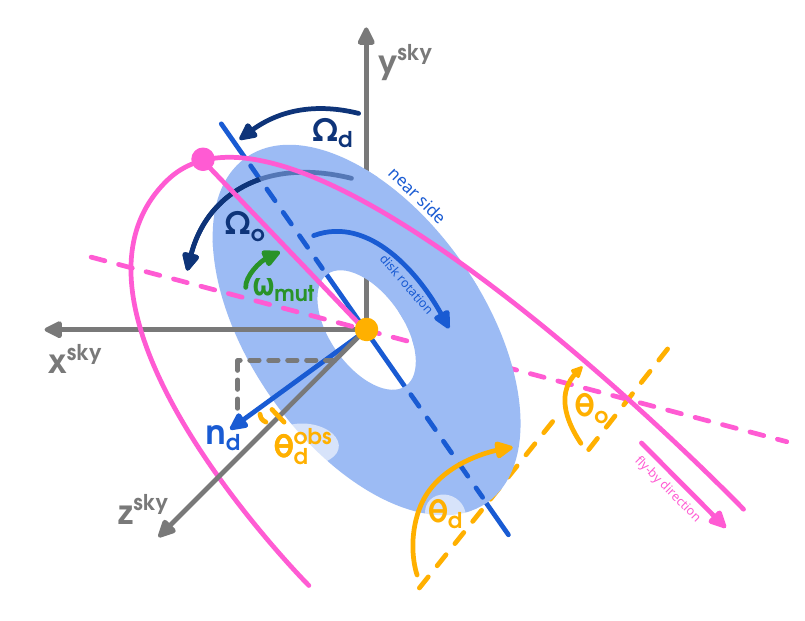}
    \caption{Geometry of the disk and fly-by orbit of the RW~Aur system in the reference frame of the sky.
    Light blue indicates the disk, pink indicates the orbit, where the pink dashed line shows the crossing line of the plane of the orbit with the $x$-$y$-plane and the periastron (pink dot) is ``in front'', meaning closer to us than the $x$-$y$-plane. The position angles $\Omega_\mathrm{d,o}$, inclinations $\theta_\mathrm{d,o}$ and the argument of the periapsis of the orbit $\omega_\mathrm{o}$ are defined consistently with our definition in Figure~\ref{fig:flyby-geometry}.
Note that this coordinate system is left-handed.
    }
    \label{fig:rw-aur-geometry}
\end{figure}

In this section, we derive the equations to find the mutual geometry of disk and orbit, in the case that the geometry of the disk plane and the orbit is given in the reference frame of the sky.
We performed this consideration on the basis of the RW~Aur geometry system, which is why we chose some definitions that are convenient for this specific case.
However, this geometry consideration can be applied to other systems if careful attention is brought on these definitions.
We define a left-handed coordinate system for the sky frame\footnote{This left-handed coordinate system is motivated by the fact that East and West are typically flipped in observational coordinate systems.}, with the $z^\mathrm{sky}$-axis pointing away along the line of sight, as shown in Figure~\ref{fig:rw-aur-geometry}.

An observed disk is characterized by a position angle $\Omega_\mathrm{d}$, which corresponds to the longitude of ascending node in the plane of the sky, and an inclination $\theta_\mathrm{d}$ with respect to the sky plane.
From these angles, we can construct the normal vector of the disk in the reference frame of the sky
\begin{equation} \label{eq:norm-vec-d}
    \vec{n_\mathrm{d}} = \begin{pmatrix}
        -\sin(\theta_\mathrm{d}) \cos(\Omega_\mathrm{d})\\
        -\sin(\theta_\mathrm{d}) \sin(\Omega_\mathrm{d})\\
        \cos(\theta_\mathrm{d})
    \end{pmatrix},
\end{equation}
where $\Omega_\mathrm{d}$ is the longitude of ascending node of the disk, corresponding to the crossing line of the disk with the plane of the sky. This angle is defined from North, or positive $y$-axis, as indicated in Figure~\ref{fig:rw-aur-geometry}.

It is important to note that according to the geometry definition of trajectory to disk in Figure~\ref{fig:flyby-geometry}, the inclination angle $\theta_\mathrm{d}$ is not the same angle as the inclination typically given in observations $\theta_\mathrm{d}^\mathrm{obs}$.
This $\theta_\mathrm{d}^\mathrm{obs}$ is defined as the angle between the normal vector and the part of the line of sight pointing toward us (see Figure~\ref{fig:rw-aur-geometry}).
By this definition, $\theta_\mathrm{d} = \pi - \theta_\mathrm{d}^\mathrm{obs}$.
When we plug this into the normal vector in Equation~\ref{eq:norm-vec-d}, with $\sin(\pi - \theta)~=~\sin(\theta)$ and $\cos(\pi - \theta)~=~-\cos(\theta)$ we see that the third component is negative for $|\theta_\mathrm{d}^\mathrm{obs}| < \pi$.
This aligns with our definition of a left-handed coordinate system.

The orbit is characterized by a position angle $\Omega_\mathrm{o}$, an inclination $\theta_\mathrm{o}$, and an argument of periapsis $\omega_\mathrm{o}$.
Here, the definition of inclination extracted from observations usually aligns with the inclination definition in Figure~\ref{fig:flyby-geometry}, which is why we can use the value from observations directly.
The normal vector of the plane can be described analogous to the disk's normal vector
\begin{equation}
    \vec{n_\mathrm{o}} = \begin{pmatrix}
        -\sin(\theta_\mathrm{o}) \cos(\Omega_\mathrm{o})\\
        -\sin(\theta_\mathrm{o}) \sin(\Omega_\mathrm{o})\\
        \cos(\theta_\mathrm{o})
        \end{pmatrix}.
\end{equation}

\subsection{Mutual inclination}
The mutual inclination between the disk plane and the orbital plane then simply corresponds to the angle between the two normal vectors $\vec{n}_\mathrm{d}$ and $\vec{n}_\mathrm{o}$.
This can be calculated by
\begin{equation}
    \cos(\theta_\mathrm{mut}) = \frac{\vec{n}_\mathrm{d} \cdot \vec{n}_\mathrm{o}}{|\vec{n}_\mathrm{d}| \ |\vec{n}_\mathrm{o}|} = \vec{n}_\mathrm{d} \cdot \vec{n}_\mathrm{o},
\end{equation}
where the last term holds true because both vectors are unit vectors.
This results in
\begin{equation}
    \cos(\theta_\mathrm{mut}) = \sin(\theta_\mathrm{d}) \sin(\theta_\mathrm{o}) \cos(\Omega_\mathrm{d} - \Omega_\mathrm{o}) + \cos(\theta_\mathrm{d}) \cos(\theta_\mathrm{o}).
\end{equation}
The same expression can for example be found in \citet{VanderPlas2019}, their Equation~4 or \citet{Gonzalez2020}, their Equation~2.

\subsection{Argument of periapsis of the orbit with respect to the disk plane}
Finding the argument of periapsis with respect to the disk plane is not trivial.
\citet{Gonzalez2020} consider a similar geometric problem in the disk of HD 100453. However, in their SPH simulations, they were able to measure this angle. Here, we aim to derive an expression to calculate the angle from the given orientations of disk and orbit on the sky.

To derive the equation for this, we construct a vector along the longitude of ascending node of the orbit along the disk plane in the frame of the sky coordinate system, and a vector pointing to the periapsis. The argument of periapsis of the orbit with respect to the disk plane (which we will call mutual argument of periapsis in the following) then corresponds to the angle between those two constructed vectors.

First, we construct the vector along the longitude of ascending node of the orbit within the disk plane in the coordinate system of the sky.
We call this vector $\vec{d}$ and calculate it with the cross product of the two normal vectors
\begin{equation}
    \vec{d}^\mathrm{sky} = \vec{n}_\mathrm{d} \times \vec{n}_\mathrm{o}    = \begin{pmatrix}
        -\sin(\theta_\mathrm{d}) \sin(\Omega_\mathrm{d}) \cos(\theta_\mathrm{o}) + \cos(\theta_\mathrm{d}) \sin(\theta_\mathrm{o}) \sin(\Omega_\mathrm{o})\\
    -\cos(\theta_\mathrm{d}) \sin(\theta_\mathrm{o}) \cos(\Omega_\mathrm{o}) + \sin(\theta_\mathrm{d}) \cos(\Omega_\mathrm{d}) \cos(\theta_\mathrm{o}) \\
    \sin(\theta_\mathrm{d}) \sin(\theta_\mathrm{o}) \sin(\Omega_\mathrm{o} - \Omega_\mathrm{d})
    \end{pmatrix}.
\end{equation}

\begin{figure}[ht!]
    \centering
    \includegraphics[width=0.4\linewidth]{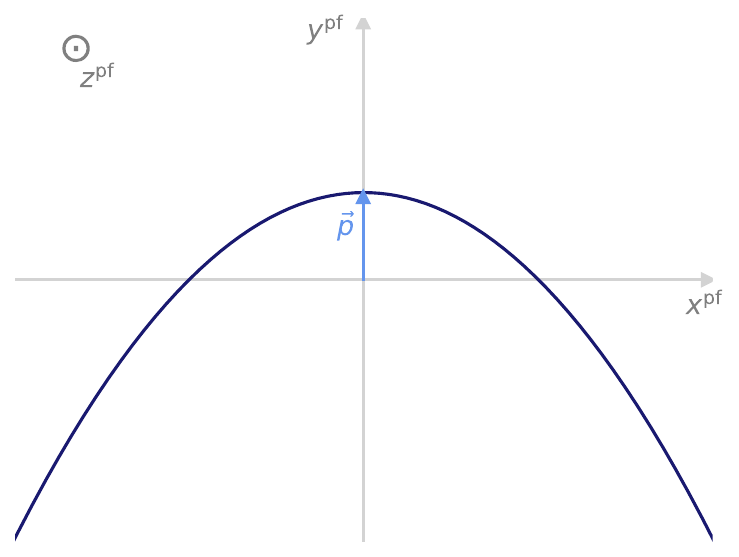}
    \caption{Our definition of the perifocal frame, which lies in the plane of the orbit.}
    \label{fig:perifocal-frame}
\end{figure}

Second, we need to construct the vector pointing to the periapsis. For this, we define a new coordinate system in the plane of the orbit. We define it such that the periapsis lies on the $y_\mathrm{pf}$-axis.
We use the index $\mathrm{pf}$ for this coordinate system, as it is sometimes called perifocal frame (see Figure~\ref{fig:perifocal-frame}).

In this perifocal frame, the coordinates of the unit vector pointing to the periapsis are trivial: $\vec{p}^\mathrm{pf}~=~(0, 1, 0)$.
To construct the coordinates of this vector in the reference frame of the sky, we need to rotate the coordinate system along all three angles that characterize the orientation of the orbit.
As these characterizing angles are defined counter-clockwise, we need to rotate the coordinate system clockwise in order to find the correct coordinates.
In a left-handed coordinate system, the following rotation matrices describe clockwise rotations
\begin{equation}
    R_\omega =
    \begin{pmatrix}
        \cos(\omega_\mathrm{o}) & -\sin(\omega_\mathrm{o}) & 0 \\
        \sin(\omega_\mathrm{o}) & \cos(\omega_\mathrm{o}) & 0 \\
        0 & 0 & 0
    \end{pmatrix}, \\
    R_\theta =
    \begin{pmatrix}
        \cos(\theta_\mathrm{o}) & 0 & \sin(\theta_\mathrm{o}) \\
        0 & 1 & 0 \\
        -\sin(\theta_\mathrm{o}) & 0 & \cos(\theta_\mathrm{o})
    \end{pmatrix}, \ \ \mathrm{and} \\
    R_\Omega =
    \begin{pmatrix}
        \cos(\Omega_\mathrm{o}) & -\sin(\Omega_\mathrm{o}) & 0 \\
        \sin(\Omega_\mathrm{o}) & \cos(\Omega_\mathrm{o}) & 0 \\
        0 & 0 & 0
    \end{pmatrix}.
\end{equation}
Using these matrices, we can determine the coordinates of $\vec{p}$ in the frame of the sky
\begin{equation}
    \vec{p}^\mathrm{sky} = R_\Omega\ R_\theta\ R_\omega\ \vec{p}^\mathrm{pf} =
    \begin{pmatrix}
        -\cos(\Omega_\mathrm{o}) \cos(\theta_\mathrm{o}) \sin(\omega_\mathrm{o}) - \sin(\Omega_\mathrm{o}) \cos(\omega_\mathrm{o})\\
        -\sin(\Omega_\mathrm{o}) \cos(\theta_\mathrm{o}) \sin(\omega_\mathrm{o}) + \cos(\Omega_\mathrm{o}) \cos(\omega_\mathrm{o})\\
        \sin(\theta_\mathrm{o}) \sin(\omega_\mathrm{o})
    \end{pmatrix}.
\end{equation}
We note that the rotations do not change the length of this vector, meaning that it still is a unit vector.
With this, we can finally calculate the angle between $\vec{d}^\mathrm{sky}$ and $\vec{p}^\mathrm{sky}$, which gives the mutual argument of periapsis
\begin{equation}
   \cos(\omega_\mathrm{mut}) =  \vec{d}^\mathrm{sky} \cdot \vec{p}^\mathrm{sky} 
     = \sin(\omega_\mathrm{o}) \sin(\theta_\mathrm{d}) \cos(2 \theta_\mathrm{o}) \sin(\Omega_\mathrm{d} - \Omega_\mathrm{o}) 
     + \cos(\omega_\mathrm{o}) \ [ \sin(\theta_\mathrm{d}) \cos(\theta_\mathrm{o}) \cos(\Omega_\mathrm{o} - \Omega_\mathrm{d})
      - \cos(\theta_\mathrm{d}) \sin(\theta_\mathrm{o}) ].
\end{equation}

\section{Further evaluation of the RW~Aur models}

\subsection{Different disk viscosities} \label{sec:appendix-rwaur-viscosities}

Because spirals might have different lifetimes for different disk viscosities, we investigate the hydrodynamical results in one simulation with a larger $\left(\alpha=10^{-2}\right)$ and one with a lower disk viscosity $\left(\alpha=10^{-4}\right)$.
All other parameters, including the trajectory of the fly-by, remain exactly the same.

\begin{figure}[ht!]
\centering
\begin{minipage}{.45\textwidth}
   \centering
    \includegraphics[width=\linewidth]{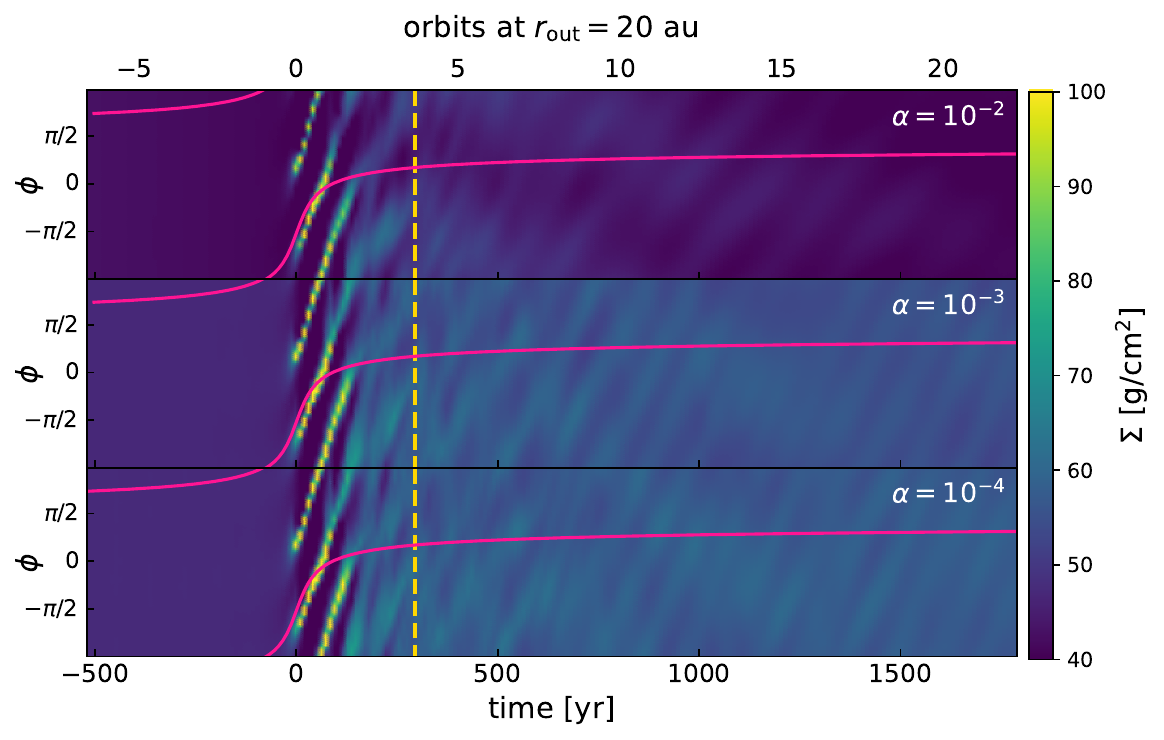}
    \caption{Azimuthal surface density profile evolution at $r=15\,\mathrm{au}$ for the simulations with different disk viscosities. The yellow dashed line indicates the current time and the pink line indicates the angle to the perturber.}
    \label{fig:rwaur-spirals-viscosity}
\end{minipage} \hspace{0.5cm}
\begin{minipage}{.45\textwidth}
    \centering
    \includegraphics[width=0.8\linewidth]{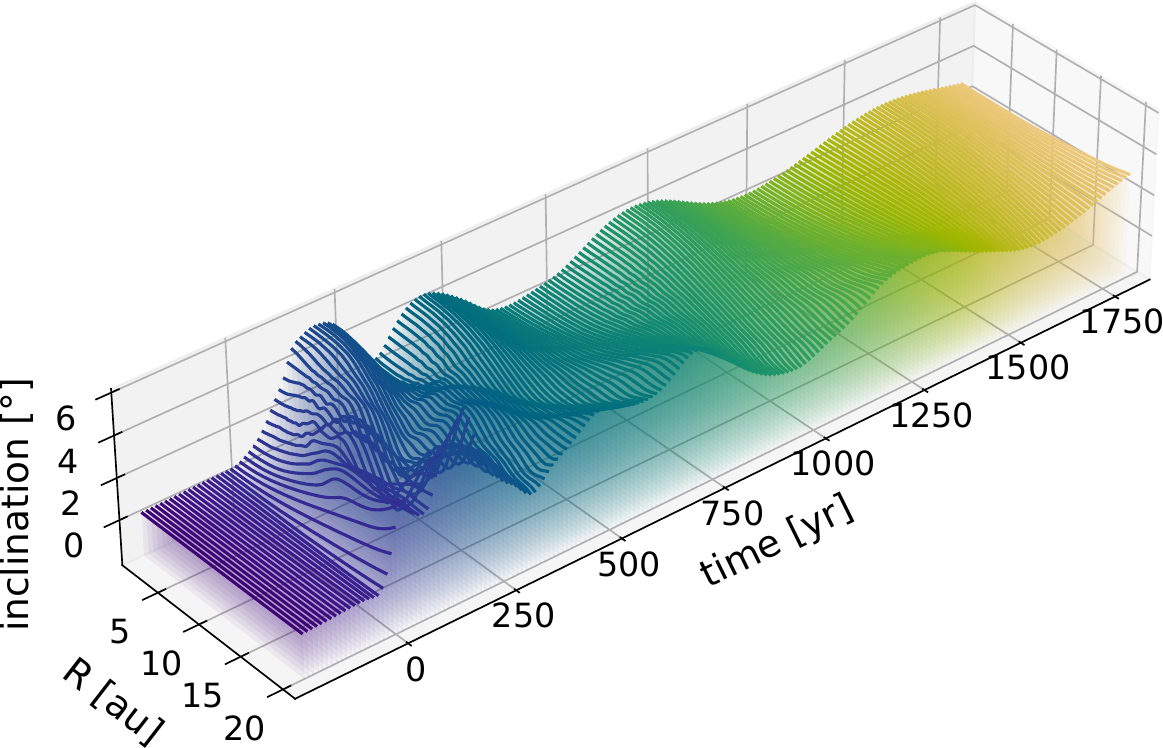}
    \includegraphics[width=0.8\linewidth]{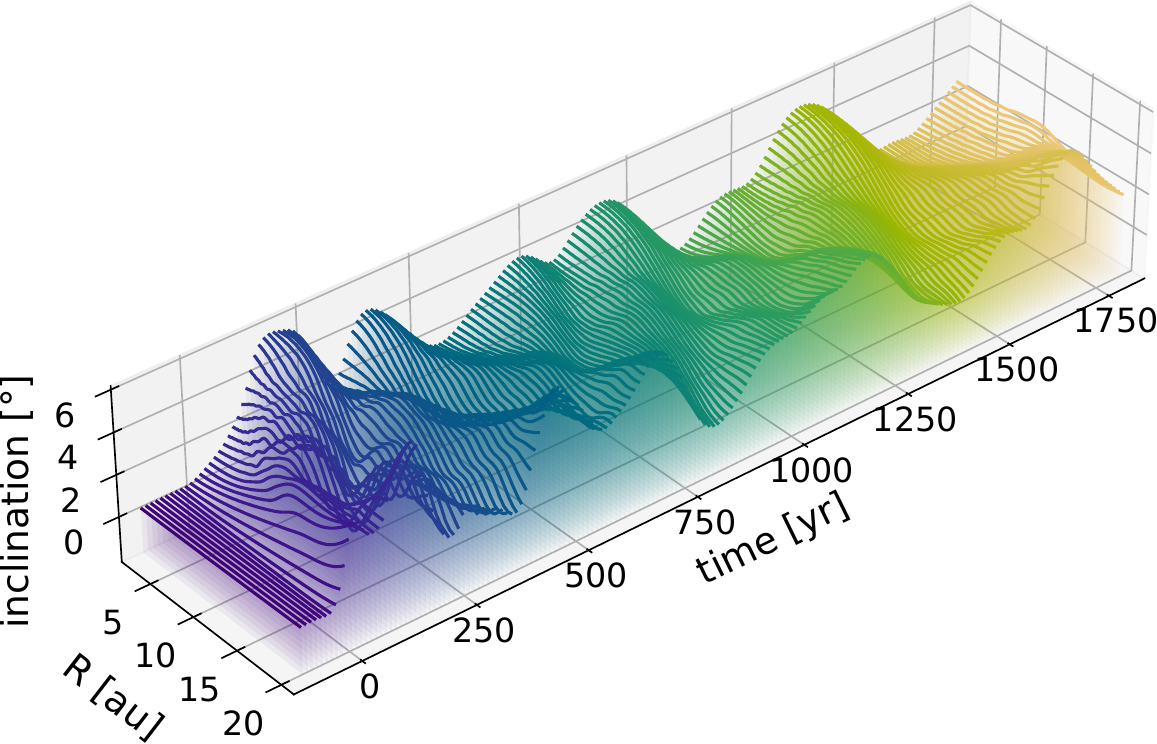}
    \caption{Evolution of the inclination profile in the simulation with ${\alpha=10^{-2}}$ (top) and ${\alpha=10^{-4}}$ (bottom).}
    \label{fig:rwaur-inclevol-viscosity}
\end{minipage}
\end{figure}

Figure~\ref{fig:rwaur-spirals-viscosity} shows the evolution of the azimuthal profile at $r=15\,\mathrm{au}$ for the simulations with different viscosities. The middle panel here shows the fiducial simulation presented before.
The lifetime of the spirals does not depend strongly on the viscosity.
As expected, the spirals dissolve slightly faster for higher viscosity.
However, for lower viscosity, no significant difference in spiral lifetime is visible.
This means that we do not expect to see strong spirals at the current time of observations, even if different assumptions on viscosity are taken.
The lifetime of the spirals could, however, be dependent on other disk properties, such as initial disk size and vertical density structure (inter alia, flaring).

Investigating the warp in these simulations in Figure~\ref{fig:rwaur-inclevol-viscosity}, we find that the amplitude of the excited warp also does not depend on the disk viscosity.
However, the viscosity influences the evolution of the warp after the fly-by. This is especially visible for the largest viscosity simulation as the warp is dampened rapidly.
The two low-viscosity simulations show no strong differences on the short simulated timescales.
However, we expect a longer lifetime of the warp for the lowest viscosity.

\subsection{Additional radiative transfer data} \label{sec:unconvolved}

\begin{figure}[ht!]
\centering
\begin{minipage}{.65\textwidth}
    \centering
    \includegraphics[width=0.9\linewidth]{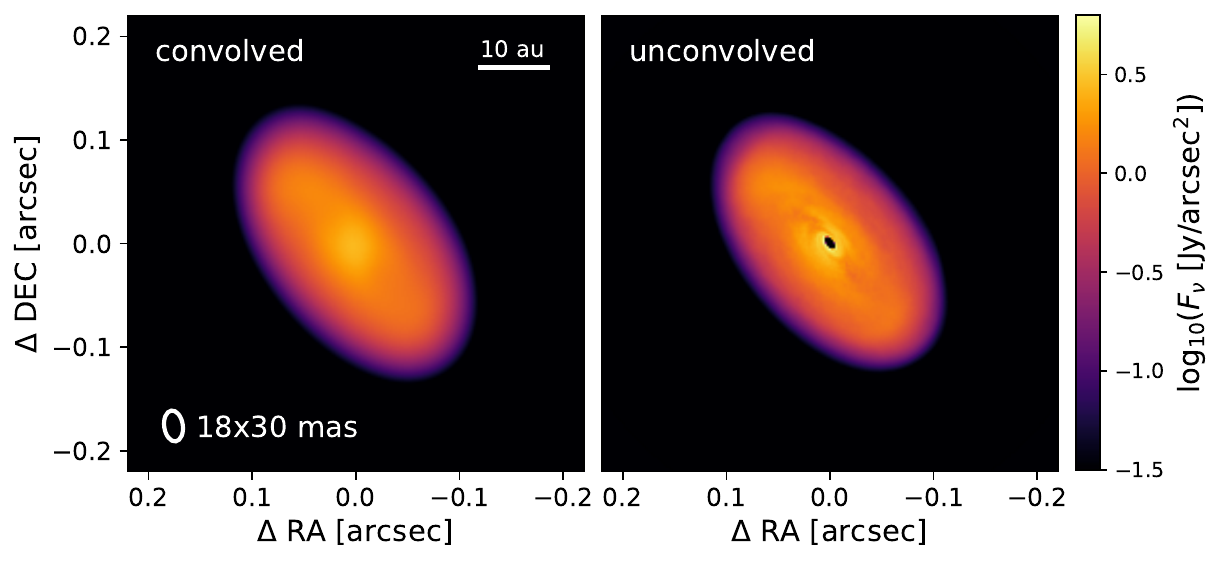}
    \caption{Like Figure~\ref{fig:rwaur-observation}, but both panels are the radiative transfer model. The left panel is convolved with a Gaussian beam, the right panel shows the unconvolved, raw output from the radiative transfer model.}
    \label{fig:unconvolved-radtrans}
\end{minipage} \hspace{0.5cm}
\begin{minipage}{.3\textwidth}
    \centering
    \includegraphics[width=\linewidth]{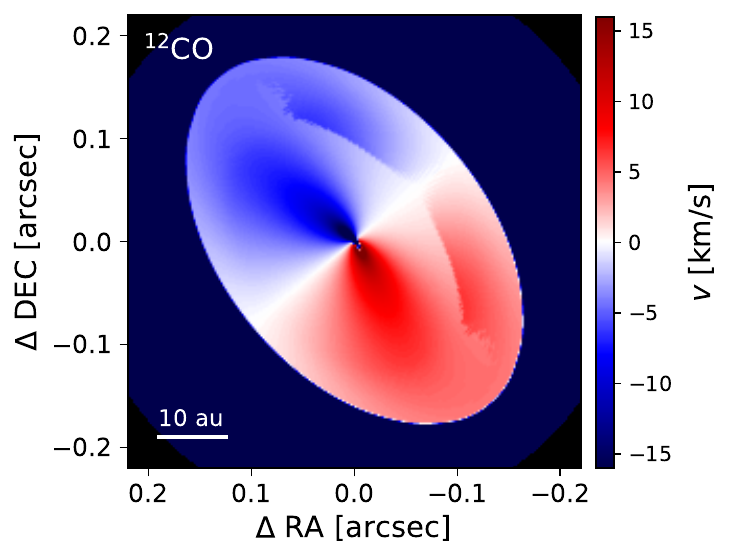}
    \caption{Velocity of the peak intensity of the initial setup of the hydrodynamic simulation of RW Aur A. We note that a part of the backside of the disk is visible at the near side of the disk in the upper right corner of the image.}
    \label{fig:rwaur-keplerian-model}
\end{minipage}
\end{figure}

Figure~\ref{fig:unconvolved-radtrans} shows the convolved and unconvolved radiative transfer simulation in comparison to each other.
In the unconvolved image, weak spiral structures close to the inner edge of the disk are visible. However, they disappear with the convolution.

In Figure~\ref{fig:rwaur-keplerian-model}, we show the velocity of the peak intensity we used to compute the residuals presented in Figure~\ref{fig:rwaur-kinematic-residuals}. We created this map using the initial setup of our hydrodynamic simulation of RW Aur A, which is a Keplerian disk model with corrections for pressure gradients.

\end{appendix}

\end{document}